\documentclass{article}

\usepackage{arxiv}
\usepackage{algorithm,algpseudocode}

\usepackage[utf8]{inputenc} 
\usepackage[T1]{fontenc}    
\usepackage{hyperref}       
\usepackage{url}            
\usepackage{booktabs}       
\usepackage{amsfonts}       
\usepackage{nicefrac}       
\usepackage{microtype}      
\usepackage{lipsum}

\usepackage{times}
\usepackage[super,sort&compress]{natbib}
\setcitestyle{comma,numbers,super,open={},close={}} 
\makeatletter
\renewcommand\@biblabel[1]{#1.}
\usepackage[]{graphicx}
\chardef\bslash=`\\ 

\newcommand{\Prob}{\ensuremath{\textrm{P}}}
\newcommand{\Exp}{\ensuremath{\textrm{E}}}

\usepackage{color}
\usepackage{hyperref}
\usepackage[T1]{fontenc}
\usepackage{amsmath}

\newcommand{\given}{\,\vert\,}

\title{Simulating and reporting frequentist operating characteristics of clinical trials that borrow external information}

\author{
Annette Kopp-Schneider \\
  Division of Biostatistics\\
  German Cancer Research Center\\
  Heidelberg, Germany \\
  \texttt{kopp@dkfz.de} \\
   \And
 Manuel Wiesenfarth \\
  Division of Biostatistics\\
  German Cancer Research Center\\
  Heidelberg, Germany \\
  \And
Leonhard Held \\
  Epidemiology, Biostatistics and\\
  Prevention Institute\\
  University of Zurich\\
  Zurich, Switzerland \\
   \And
 Silvia Calderazzo \\
  Division of Biostatistics\\
  German Cancer Research Center\\
  Heidelberg, Germany \\
}

\begin{document}
\maketitle
\begin{abstract}
Borrowing of information from historical or external data to inform inference in a current trial is an expanding field in the era of precision medicine, where trials are often performed in small patient cohorts for practical or ethical reasons. Many approaches for borrowing from external data have been proposed. Even though these methods are mainly based on Bayesian approaches by incorporating external information into the prior for the current analysis, frequentist operating characteristics of the analysis strategy are of interest. In particular, type I error and power at a prespecified point alternative are in the focus. It is well-known that borrowing from external information may lead to the alteration of type I error rate.  
We propose a procedure to investigate and report the frequentist operating characteristics in this context. The approach evaluates type I error rate of the test with borrowing from external data and calibrates the test without borrowing to this type I error rate.  On this basis, a fair comparison of power between the test with and without borrowing is achieved.
\end{abstract}


 \textcolor{red}{This paper has not been peer reviewed yet.} 

\maketitle  

\section{Introduction}
When trials can only be performed with small sample sizes as, e.g., in the situation of precision medicine where patients cohorts are defined by a specific combination of biomarker and targeted therapy, borrowing of information from historical data is currently discussed as an approach to improve the efficiency of the trial. In this context, borrowing of information is often also referred to as evidence synthesis or extrapolation, where external data could be historical data or another source of co-data.
The number of approaches for borrowing from external data that dynamically discount the amount of information transferred from external data based on the discrepancy between the external and current data is increasing steadily. Even though these methods are mainly based on Bayesian approaches by incorporating external information into the prior for the current analysis, frequentist operating characteristics of the analysis strategy are of interest. In particular, type I error (T1E) and power at a prespecified point alternative are in the focus.  

Frequentist operating characteristics of borrowing methods are often investigated by using simulation studies. The EMA states that "a common approach to addressing the risk of type-I error rate inflation when information is borrowed is to carry out multiple simulation studies to quantify this effect." \cite{EMAProcova}.
The aim of such a simulation study is to provide an overall assessment of the performance of the borrowing method. 

The actual performance of a method for borrowing from external information depends on the specific external information that is borrowed from, i.e., it is derived for a fixed constellation of external data. When the overall performance of a borrowing method should reflect its operating characteristics for yet unknown constellations of external data, simulation studies have been performed in which varying external data are simulated.
The aim of this work is to discuss how frequentist operating characteristics of borrowing methods should be simulated and to suggest how to quantify and report power in relationship to T1E rate. Since the relationship between T1E rate and power of a test is non-linear and not intuitive, our approach is to evaluate T1E rate of the test with borrowing from external data, and to compare the power of the test with borrowing to the power of the test without borrowing at this T1E rate, and to thus allow for a fair comparison of the power of the test with and without borrowing. Often, external data are known at the planning stage of a trial and hence can be considered as fixed. However, there are circumstances in which external data should be considered as random. We address both constellations, provide algorithms for assessing frequentist operating characteristics and show the difference in results. 

The problem setup discussing hypothesis testing in the frequentist and the Bayesian approach is described in Section \ref{Problem}. Section \ref{fixed} proposes an algorithm to simulate and illustrate potential power gains/losses when borrowing from external data, considering external data as fixed and known when frequentist operating characteristics are derived. The algorithm is illustrated for the situation of the one-arm one-sided normal test and for the two-arm one-sided normal test with borrowing to the control arm, i.e. the hybrid control arm trial situation. Section \ref{random external} deals with the situation that external data is considered as random. An algorithm is proposed as well and applied for the situations investigated in Section \ref{fixed}. We discuss our findings in Section \ref{Discussion}.

\section{Problem setup}\label{Problem}
Similar to the notation in  \cite{kopp-schneider2020} and in \cite{Lehmann1986}, assume that the endpoint has probability density function $f_\theta$ and the hypotheses investigated are
\begin{equation}
\nonumber
H_0: \theta \in \Theta_0 \text{ vs. } H_1: \theta \notin \Theta_0. 
\end{equation}
Typically, the hypotheses are one-sided, i.e., without loss of generality,
\begin{equation}
\nonumber
H_0: \theta \leq \theta_0 \text{ vs. } H_1: \theta > \theta_0. 
\end{equation}

Let $D= \{D_1,\ldots,D_n\} $ be the random variables from which the observations of the current trial, $d= \{d_1,\ldots,d_n\} $, are obtained. Note that capital letters indicate random variables whereas lowercase letters indicate the observations. For trial evaluation, the test decision will be performed by the test
\begin{equation}
\varphi^{(\alpha)}(d) = 
\begin{cases}
1 & \text{if } T(d) \in \mathcal{C} \\
0 & \text{if } T(d) \notin \mathcal{C} 
\end{cases}
\end{equation}
where $T(d_1,\ldots,d_n)=T(d)$ is a sufficient test statistic for $f_\theta$ and the rejection region $\mathcal{C}$ is chosen as the largest set such that $\underset{\theta \in \Theta_0}{\max} \{ \Exp_{\theta}[\varphi^{(\alpha)}(D)]\}\leq \alpha$, i.e., the test controls (frequentist) T1E rate, $\alpha$ denoting the significance level of the test.
While for continuous distributions $f_\theta$, $\mathcal{C}$ can be selected such that $\underset{\theta \in \Theta_0}{\max} \{ \Exp_{\theta}[\varphi^{(\alpha)}(D)]\}= \alpha$, the significance level may not be attained for discrete distributions. For a given data set $d$, the value of the test function $\varphi^{(\alpha)}(d)$ corresponds to the probability to reject $H_0$. 
In case that a uniformly most powerful (UMP) test exists for the test situation, the rejection region $\mathcal{C}$ is an interval $(t_0,\infty)$ with appropriately chosen $t_0$. 

Now assume that external information $d_E=\{d_{E,1},\ldots,d_{E,n_E}\}$, independent of $d$, is available and should be used to inform the test decision for the current trial. 
A decision rule is formulated based on the observed results, $d$, of the current trial and the external information $d_E$ which is a fixed observation as well.
Accordingly, a test function $\varphi_{\text{B}}$ (the index ``B'' indicating that it is a test with borrowing) is identified with potentially different rejection region $\mathcal{C}_{d_E}$ such that 
\begin{equation}
\label{Eq:phiB}
\varphi_{\text{B}}(d;d_E) = 
\begin{cases}
1 & \text{if } T(d) \in \mathcal{C}_{d_E}  \\
0 & \text{if } T(d) \notin \mathcal{C}_{d_E}
\end{cases}
\end{equation}
As the external information $d_E$ is fixed, $\varphi_{\text{B}}(.;d_E)$ is in fact a function only of the current data $d$.
The T1E rate $\underset{\theta \in \Theta_0}{\max} \{ \Exp_{\theta}[\varphi_{\text{B}}(D;d_E)]\}=\underset{\theta \in \Theta_0}{\max} \{ \Exp_{\theta}[\varphi_{\text{B}}(D;D_E)\given D_E=d_E]\}$ may differ from $\alpha$. For the following we use the notation $\alpha_{\text{B}}(\theta;d_E)= \Exp_{\theta}[\varphi_{\text{B}}(D;D_E)\given D_E=d_E]$ for $\theta \in H_0$ and $\alpha_{\text{B}}(d_E)=\underset{\theta \in \Theta_0}{\max} \{\alpha_{\text{B}}(\theta;d_E)\}$.

Even though testing was introduced here in the frequentist framework, the test $\varphi^{(\alpha)}$ can also be formulated in a Bayesian framework based on the posterior probability $\Prob(\theta > \theta_0 \given  D=d)$ such that \cite{held2020}
\begin{equation}
\label{Eq:phiBayes}
\varphi^{(\alpha)}(d) = 
\begin{cases}
1 & \text{if } \Prob(\theta > \theta_0 \given  D=d) > c \\
0 & \text{if } \Prob(\theta > \theta_0 \given  D=d) \leq c
\end{cases}
\end{equation}
with a selected (non-informative) prior and appropriately chosen $c \in [0,1)$ such that T1E rate is controlled at $\alpha$. Our approach is hence based on tail probabilities rather than Bayes factors, the more principled Bayesian approach to hypothesis testing. 
As argued in \cite{kopp-schneider2020}, the decision function $\varphi_{\text{B}}(D;d_E)$ for the one-sided test is accordingly given by 
\begin{equation}
\label{Eq:phiBBayes}
\varphi_{\text{B}}(d;d_E) = 
\begin{cases}
1 & \text{if } \Prob(\theta > \theta_0 \given  D=d; D_E=d_E) > c_{d_E} \\
0 & \text{if } \Prob(\theta > \theta_0 \given  D=d; D_E=d_E) \leq c_{d_E}
\end{cases}
\end{equation}
where the posterior is induced by a prior that incorporates the external information $d_E$ (this is indicated here by ``$\Prob(.\given .;d_E)$''), and the threshold $c_{d_E} \in [0,1)$ . 
In case that the threshold with borrowing, $c_{d_E}$, is kept at the threshold without borrowing, $c$, the level of the test with borrowing $\alpha_{\text{B}}(d_E)$ is likely different from the original $\alpha$-level. If it exceeds $\alpha$, this is called T1E inflation due to borrowing from external data. 
Often, but not always, the threshold with borrowing, $c_{d_E}$, can be adjusted such that T1E rate is still controlled, as shown in \cite{kopp-schneider2020}. 

\section{Borrowing from fixed external data}\label{fixed}

When external data are used to augment the data set from a current trial, the external data are fixed and known already in the planning stage of the trial. The frequentist operating characteristics of the trial are then derived upfront. 
When investigating the properties of a borrowing method, one will typically not only want to describe its frequentist operating characteristics conditional on a specific external data set $d_E$, but it will be important to characterize the method in a more general context for varying external data. Hence a general approach is needed for reporting the operating characteristics of a borrowing method, also to be able to compare different borrowing methods.  

If $\alpha_{\text{B}}(d_E)$ is different from $\alpha$, a comparison of the power of $\varphi_{\text{B}}$ to the power of $\varphi^{(\alpha)}$ is difficult to interpret. 
A fair and interpretable comparison of the power of a test with and a test without borrowing should be performed in case that both tests have the same T1E rate. 
It is, however, not always possible to adjust the rejection region of a test with borrowing, $\varphi_{\text{B}}$, such that T1E rate is controlled at the original $\alpha$-level. Instead, we propose to evaluate $\alpha_{\text{B}}(d_E)$ and compare the performance of the test with borrowing to a calibrated test without borrowing at the $\alpha_{\text{B}}(d_E)$-level, i.e., we compare the performance of $\varphi_{\text{B}}(D;d_E)$ to $\varphi^{(\alpha_{\text{B}}(d_E))}(D)$. The test without borrowing at the adjusted T1E rate, $\varphi^{(\alpha_{\text{B}}(d_E))}$, will be called $\emph{test calibrated to borrowing}$.

\subsection{Recommended approach to investigating and reporting frequentist operating characteristics}\label{fixed external}

To characterize the operating characteristics of a test with borrowing in a more general context, repeated external datasets $d_E$ are generated and the performance of the test when borrowing from these is evaluated. Usually, the performance should be investigated for a specific parameter $\theta_E$ of the data generating process for the external data $D_E$. We recommend the  procedure in Algorithm \ref{alg:fixed} to provide a fair comparison between the performance of a test without and with borrowing, when evaluating power at $\theta_1 \in \Theta_1$.

\begin{algorithm}
\caption{Procedure for fixed external data}
\begin{algorithmic}
\State(F1) Select $\theta_E$ as parameter for external data.
\State(F2) Select $\theta_1 \notin \Theta_0$ for evaluating the power, i.e., the probability to reject $H_0$ if $D \sim f_{\theta_1} $.
\State(F3) Repeat for a sufficient number of times $n_{\text{sim}}$:
\begin{itemize}
\item[] (F3.a) Generate one external data set $d_E$ from $D_E \sim f_{\theta_E} $.
\item[] (F3.b) Determine T1E rate $\alpha_{\text{B}}(d_E)=\underset{\theta \in \Theta_0}{\max} \{\Exp_{\theta}[\varphi_{\text{B}}(D;d_E)]\}$ with borrowing from $d_E$.
\item[] (F3.c) Determine power 
\begin{itemize}
\item[]  - with borrowing, $\Exp_{\theta_1}[\varphi_{\text{B}}(D;d_E)]$.
\item[]  - without borrowing of the calibrated test, $\Exp_{\theta_1}[\varphi^{(\alpha_{\text{B}}(d_E))}(D)]$.
\end{itemize}
\item[] (F3.d) Record $\alpha_{\text{B}}(d_E)$ and the power difference $\Exp_{\theta_1}[\varphi_{\text{B}}(D;d_E)] - \Exp_{\theta_1}[\varphi^{(\alpha_{\text{B}}(d_E))}(D)]$.
\end{itemize}

\State(F4) Average $\alpha_{\text{B}}(d_E)$ and power difference.
\end{algorithmic}
\label{alg:fixed}
\end{algorithm}

Note that if the expectations $\Exp_{\theta}[...]$ in F3.b-c of Algorithm \ref{alg:fixed} cannot be determined analytically, they can be calculated numerically by evaluating test decisions on data simulated according to the data generating process.
The $n_{\text{sim}}$ generated pairs of $\alpha_{\text{B}}(d_E)$ and the power difference $\Exp_{\theta_1}[\varphi_{\text{B}}(D;d_E)] - \Exp_{\theta_1}[\varphi^{(\alpha_{\text{B}}(d_E))}(D)]$ can be summarized or shown graphically, see Section \ref{one-sided one-arm}. A related approach for investigating the operating characteristics of a test with borrowing in a more general context is to deterministically select the mean of the external data $\bar{d_E}$, e.g., on a grid of values, and perform steps (F3.b), F(3.c) and (F3.d). To illustrate the procedure, we apply it in two common situations occurring in clinical trials in the remainder of this section. 

\subsection{One-sided one-arm normal test}\label{one-sided one-arm}

For illustration, we use a simple example by assuming a one-arm trial with normally distributed endpoint with known variance, i.e., we assume that current data $D_i \sim N(\theta, \sigma^2), i=1,...,n$, and external data $D_{E,j} \sim N(\theta_E, \sigma_E^2), j=1,...,n_E$,  and we test $H_0: \theta \leq \theta_0 = 0 \text{ vs. } H_1: \theta > 0$. Assume $ \sigma=\sigma_E=1$ and sample sizes for current and external data, $n=25$ and $n_E=20$. With this setup, the normal test to level $\alpha=0.025$ has power $\Exp_{\theta_1}[\varphi^{(\alpha)}(D)]=0.705$ at $\theta_1=0.5 \in H_1$. The one-sided one-sample normal test is a UMP test. Hence it is clear \cite{kopp-schneider2020} that no borrowing method can lead to a power gain compared to a test calibrated to borrowing. 

\subsubsection{Borrowing from external data using a fixed power prior approach}\label{fixed PP}

A common borrowing approach is the use of a power prior in which the prior for the current data is obtained as a discounted posterior of the external data analysis. The downweight of external information is obtained by raising the external data likelihood $L(\theta; d_E)$ to a power $\delta \in [0,1]$ \cite{ibra2000}. The weight parameter $\delta$ can be fixed \textit{a priori} or in turn estimated from the data, in the latter case the prior has the form \cite{duan2006,neuenschwander2009note}
\begin{equation}
\nonumber
\pi(\theta,\delta\given d_E)= \frac{L(\theta; d_E)^\delta \pi(\theta) \pi(\delta)}{c(\delta)},
\end{equation}
where $c(\delta)$ is the normalizing constant $c(\delta)=\int L(\theta; d_E)^\delta \pi(\theta) d \theta$.

The weight parameter determines how much of the external information is incorporated. Extreme cases are $\delta=0$, when information from $d_E$ is discarded and $\delta=1$, when $d_E$ is completely taken into account. Note that, for normal outcomes and fixed $\delta$, there is a direct correspondence between $\delta$ and the heterogeneity parameter of two additional popular borrowing approaches, i.e., the meta-analytic \cite{neuenschwander2010} and the commensurate \cite{Hobbs2012} prior. The three approaches can indeed be shown to be equivalent for specific parameter choices \cite{chen2006,Hobbs2012}. The possibility to analytically relate the power and meta-analytic approaches in case of unknown $\delta$ and heterogeneity parameter has also been recently shown \cite{pawel2022} .

When borrowing from $d_E$ with a fixed $\delta$ and flat initial prior $\pi(\theta)$, the posterior distribution of $\theta$ is given by 

\begin{equation}
\nonumber
\pi(\theta\given d;d_E,\delta)\sim N\left(\frac{ \delta n_E \bar{d_E} + n \bar{d} }{\delta n_E + n},\frac{\sigma^2}{\delta n_E + n} \right).
\end{equation}
 where $\bar{d}$ and $\bar{d_E}$ denote the means of current and external data, resp.
Since the posterior probability $\Prob(\theta > 0 \given d;d_E,\delta)$ is monotone in the external and also in the current data mean \cite{whitt1979} 
 and following the arguments in \cite{kopp-schneider2020}, $\varphi_{\text{B}}(D;d_E)$ is a UMP test to level $\alpha_{\text{B}}(d_E)$ and hence the power of $\varphi_{\text{B}}(D;d_E)$ and the power of the test calibrated to borrowing $\varphi^{(\alpha_{\text{B}}(d_E))}$ are identical. 

\begin{figure}[ht!]
\begin{center}

\includegraphics[scale=0.7]{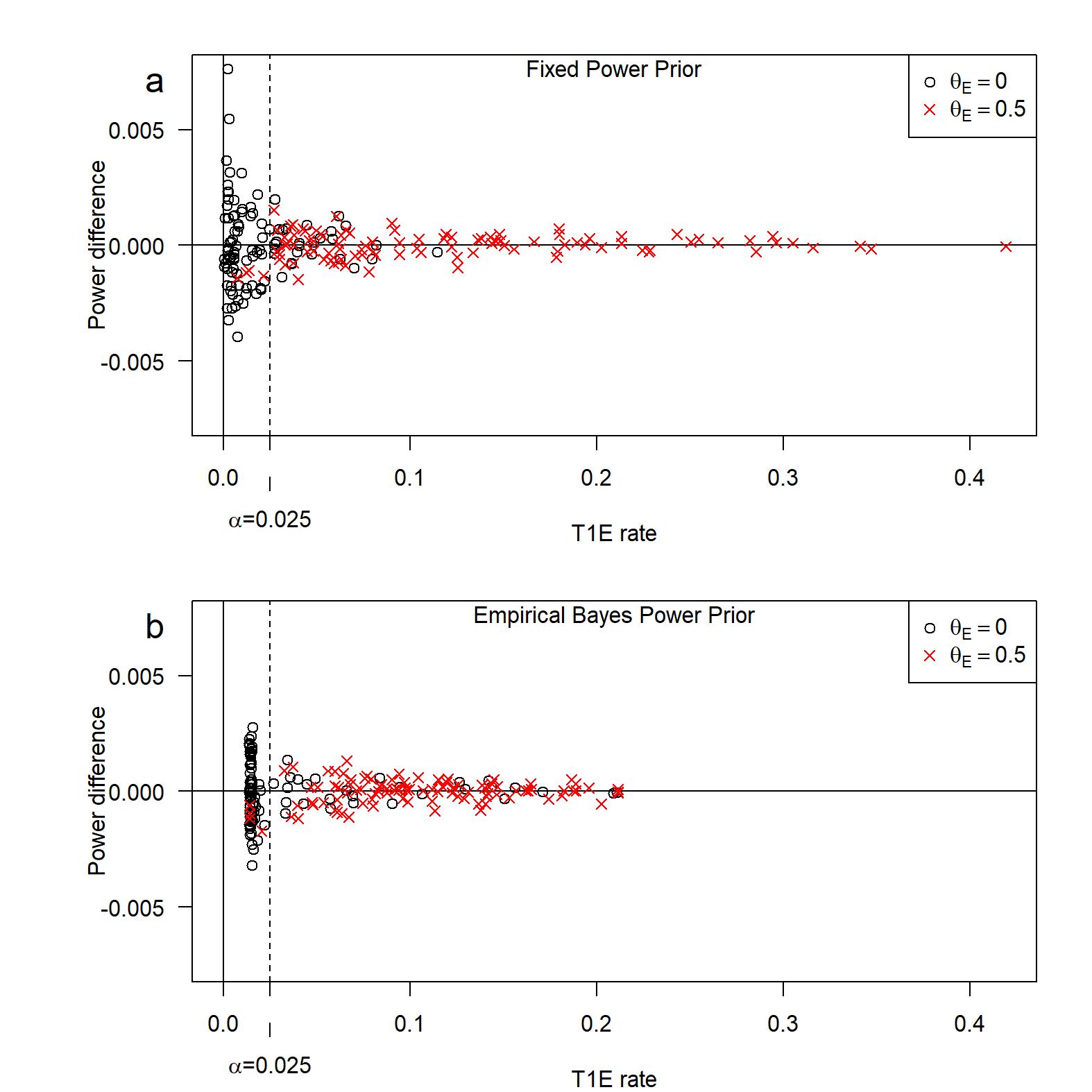}
\caption{Simulated T1E rate $\alpha_{\text{B}}(d_E)$ and power difference $\Exp_{\theta}[\varphi_{\text{B}}(D;d_E)] - \Exp_{\theta}[\varphi^{(\alpha_{\text{B}}(d_E))}(D)]$ when borrowing from external data from 100 data sets $D_E \sim N(0, 1)$ (circles) and 100 data sets $D_E \sim N(0.5, 1)$ (crosses) in the one-arm situation. Sample size for current data is $n=20$ and for external data $n_E=10$. (a) Borrowing with fixed $\delta=0.5$; (b): Borrowing with Empirical Bayes power prior. \label{fig:OneSample}}
\end{center}
\end{figure}

This situation is hence well suited for illustrating our proposed Algorithm 1. External data $d_E$ was simulated with $\theta_E=0$, i.e., on the border of $\Theta_0$, and with $\theta_E=0.5$ in the alternative, each 100 times. The weight parameter $\delta$ was set to $0.5$. A million Monte Carlo simulations were used to derive T1E rate and power with borrowing. For external data sets $d_E$ generated with $\theta_E=0$, $\alpha_{\text{B}}(d_E)$ ranges from $0.0003$ to $0.115$, observed median and mean are $0.010$ and $0.019$. For external data sets $d_E$ generated with $\theta_E=0.5$, $\alpha_{\text{B}}(d_E)$ ranges from $0.007$ to $0.419$, with median and mean $0.090$ and $0.121$. In both situations, mean power differences, which in theory should be exactly $0$, are in absolute value smaller than $10^{-4}$. 
Fig. \ref{fig:OneSample}(a) shows the power difference versus $\alpha_{\text{B}}(d_E)$ and reflects inaccuracy of the integration when calculating the power difference. Inaccuracies occur since calculation of power with borrowing requires Monte Carlo integration, and calculation of power of the test calibrated to borrowing uses $\alpha_{\text{B}}(d_E)$, which is associated with numerical inaccuracy as well.  

The operating characteristics for the external data mean $\bar{d_E}$ varying over a range of values are shown in Fig. \ref{fig:OneSample_grid} (a). In this plot scale, the power difference is exactly on the line $y=0$. T1E rate $\alpha_{\text{B}}(d_E))$ increases to $1$ as $\bar{d_E}$ increases in size, i.e. external data are located far in the alternative.

\subsubsection{Borrowing from external data using an Empirical Bayes power prior approach}
The power prior approach can be modified to adapt the power prior parameter $\delta$ to the similarity of the current and the external data, such that $\delta$ is large when the external and current data are similar and small if they are conflicting. 
We follow \citet{Gravestock2017} who propose to use an Empirical Bayes approach for estimation of $\delta(d;d_E)$ which maximizes the marginal likelihood of $\delta$, as derived in \citet{Gravestock2017}, 
\begin{equation}
\nonumber
\hat{\delta}(d;d_E)=\frac{\sigma_E^2/n}{\max\{(\bar d - \bar{d_E})^2,\sigma^2/n +\sigma_E^2/n_E\}-\sigma^2/n}.
\end{equation}
  
\begin{figure}[ht!]
\begin{center}

\includegraphics[scale=0.7]{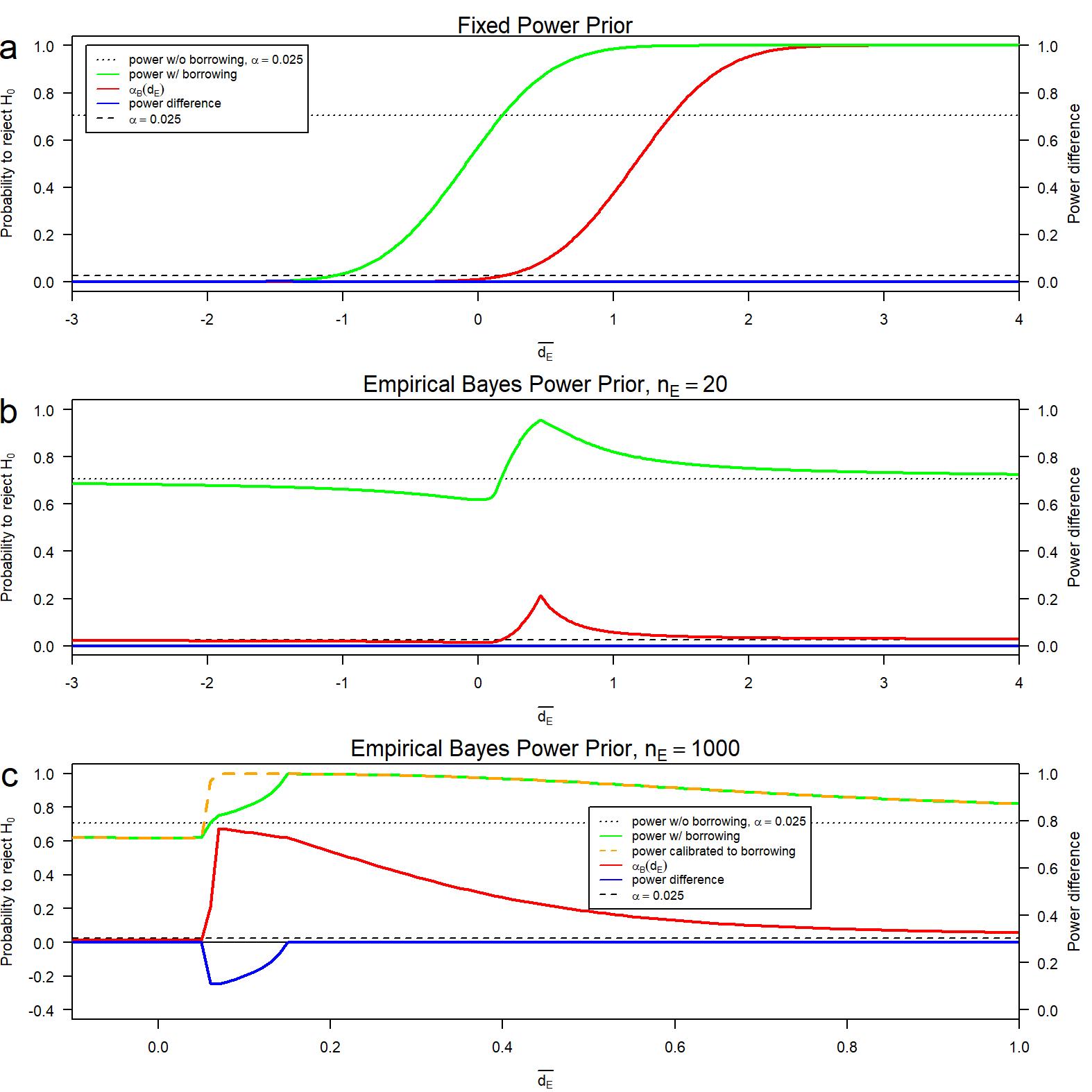}
\caption{T1E rate with borrowing ($\alpha_{\text{B}}(d_E)$), power with borrowing and power difference $\Exp_{\theta_1}[\varphi_{\text{B}}(D;d_E)] - \Exp_{\theta_1}[\varphi^{(\alpha_{\text{B}}(d_E))}(D)]$ when borrowing from external data mean for varying $\bar{d_E}$ in the one-arm situation, calculated by performing steps (F3.b) to (F3.d) in Algorithm \ref{alg:fixed}. Sample size for current data is $n=25$. (a) Borrowing with fixed $\delta=0.5$ and external data sample size $n_E=20$; (b): Borrowing with Empirical Bayes power prior and external data sample size $n_E=20$; (c): Borrowing with Empirical Bayes power prior and external data sample size $n_E=1000$. \label{fig:OneSample_grid}}
\end{center}
\end{figure}

The same external data sets $d_E$ were generated as in \ref{fixed PP}. With $\theta_E=0$, $\alpha_{\text{B}}(d_E)$ ranges from $0.014$ to $0.211$, observed median and mean are $0.015$ and $0.036$. For external data sets $d_E$ generated with $\theta_E=0.5$, $\alpha_{\text{B}}(d_E)$ ranges from $0.014$ to $0.212$, with median and mean $0.098$ and $0.105$. In both situations, mean power differences is in absolute value smaller than $10^{-4}$. Fig. \ref{fig:OneSample}(b) shows the power difference versus $\alpha_{\text{B}}(d_E)$.
The operating characteristics for the external data mean $\bar{d_E}$ varying over a range of values are shown in Fig. \ref{fig:OneSample_grid} (b). The power difference is always on the line $y=0$ in this plot scale and maximal T1E rate inflation is observed at $\bar{d_E}=0.56$.

To illustrate that dynamic borrowing can lead to a test that is not UMP, and hence that power loss can be the result of borrowing, we consider an extreme and artificial setting: we again used the Empirical Bayes power prior approach but increased the sample size of the external data to $n_E=1000$ instead of $20$. The operating characteristics for the external data mean $\bar{d_E}$ varying over a range of values are shown in Fig. \ref{fig:OneSample_grid} (c). For $0.06 \leq \bar d_E \leq 0.14$ the power of the test with borrowing is reduced compared to the test calibrated to borrowing. This is due to the fact that the posterior probability $\Prob(\theta > \theta_0 \given  D=d; D_E = d_E)$ is non-monotone in $\bar d$ and hence the rejection region $\{d\given \Prob(\theta > \theta_0 \given  D=d; D_E = d_E) > 0.975\}$ consists of two disjoint intervals. A detailed illustration of this is given in the Supplement.

\subsection{One-sided two-arm normal test with borrowing to control arm}\label{hybrid control trial}
For the case that current data from a treatment and a control trial arm are collected, we will consider the situation of a hybrid control arm trial where external (historical) data are available to inform the control arm evaluation. Also in this situation a UMP test is available and power gains are not possible when T1E rate is controlled. 

Let current control $D_c \sim N(\theta_c, \sigma^2) $, current treatment $D_t \sim N(\theta_t, \sigma^2) $, i.e. $D=(D_c,D_t)$, and external data $D_E \sim N(\theta_E, \sigma^2) $. The aim of the trial is to test 
\begin{equation}
\nonumber
H_0: \theta_t \leq \theta_c \text{ vs. } H_1: \theta_t > \theta_c
\end{equation}

or equivalently 
\begin{equation}
\nonumber
H_0: \theta_t - \theta_c \leq 0 \text{ vs. } H_1: \theta_t -\theta_c > 0.
\end{equation}

Similar to the one-sample case, the T1E rate is observed at the border of $H_0$. In contrast to the one-sample case, however, this is not just one point but a line in the two-dimensional space given by $\theta_t = \theta_c$. The T1E rate for the test without borrowing is identical for all $\theta_t = \theta_c$ independent of the actual value of $\theta_c$. In the hybrid control trial that borrows only to the current control, the T1E rate with borrowing from $d_E$ depends on $\theta_c$ as well. Determination of $\underset{\theta \in \Theta_0}{\max} \{\Exp_{\theta}[\varphi_{\text{B}}(D;d_E)]\}$ hence involves maximizing $\Exp_{\theta_t=\theta_c}[\varphi_{\text{B}}(D;d_E)]$ for all $\theta_c$ in step (F3.b). 
In step (F3.c), power is calculated for a selected $\theta_1 = \theta_t - \theta_c$. Power at $\theta_1 = \theta_t - \theta_c$ of the test without borrowing is independent of $\theta_c$. For the test with borrowing, $\Exp_{\theta_1 = \theta_t - \theta_c}[\varphi_{\text{B}}(D;d_E)]$ varies with $\theta_c$. In summary, the operating characteristics of the test with borrowing depend on three parameters $\theta_c, \theta_t$ and $d_E$, whereas it only depends on $\theta_t - \theta_c$ for the test without borrowing.

To illustrate the situation, we consider in the following an example with equal sample sizes in the current control and treatment arm, $n_c=n_t=15$, and with borrowing from $n_E=10$ external control observations. Assume $\sigma=1$ for all data. The one-sided test without borrowing is performed at level $\alpha=0.025$, power of the two-sample test without borrowing is $0.78$ for $\theta_t - \theta_c= \theta_1=1$. 

\subsubsection{Fixed power prior approach}\label{Hybrid fixed PP}
\begin{figure}[ht!]
\begin{center}

\includegraphics[scale=0.7]{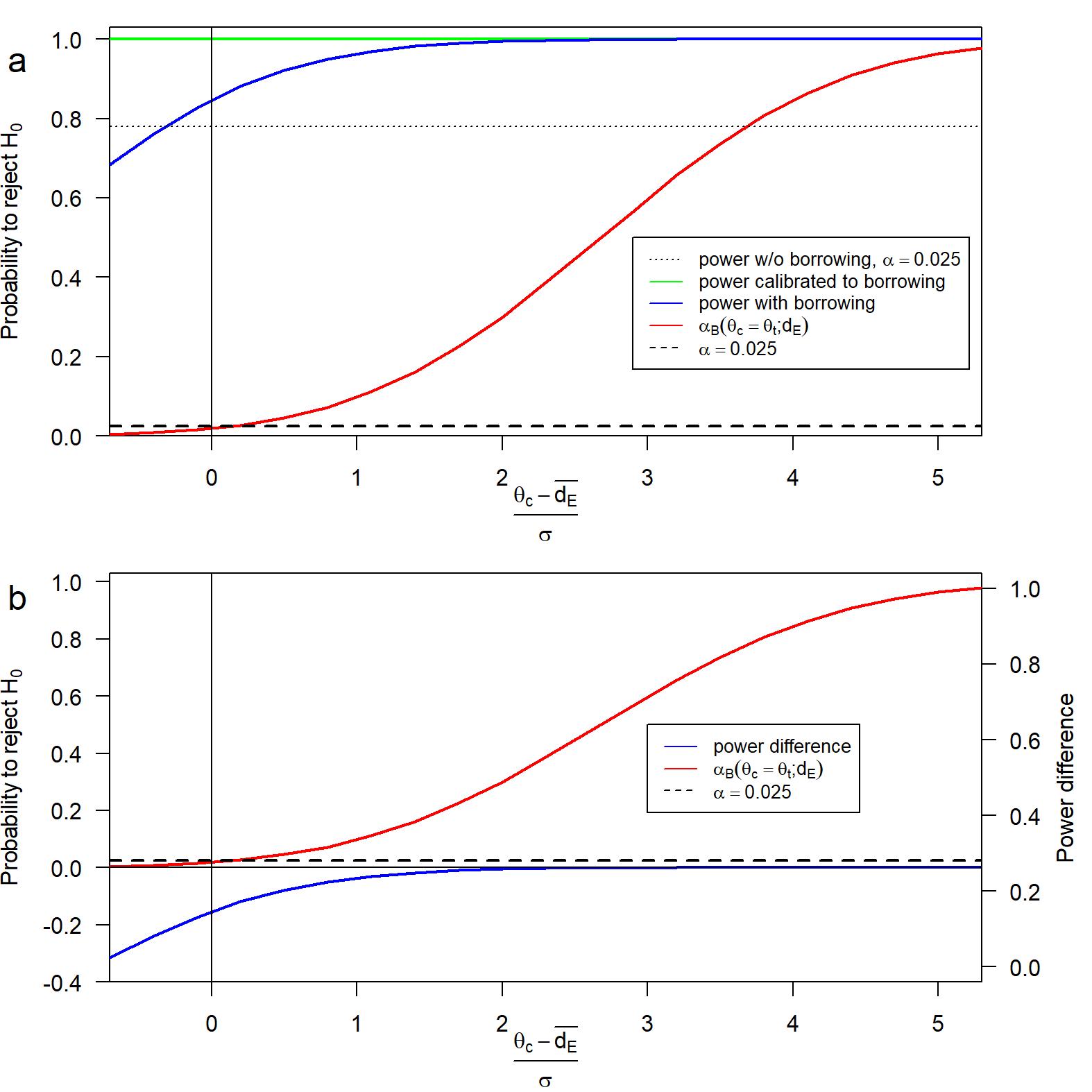}
\caption{Two-arm trial with fixed power prior borrowing of external to current control, with $\delta=0.5$. (a) $H_0$ rejection probabilities in the null hypothesis $\theta_c=\theta_t$, $\alpha_{\text{B}}(\theta_c=\theta_t;d_E)$, and on the alternative $\theta_t-\theta_c=1$ for $\varphi_{\text{B}}$ and $\varphi^{(\alpha_{\text{B}}(d_E))}$; (b) $\alpha_{\text{B}}(\theta_c=\theta_t;d_E)$ and power difference $\Exp_{\theta_t-\theta_c=1}[\varphi_{\text{B}}(D;d_E)]- \Exp_{\theta_t-\theta_c=1}[\varphi^{(\alpha_{\text{B}}(d_E))}(D)]$.  \label{fig:TwoSampleFix}}
\end{center}
\end{figure}

For illustration purposes we start with a fixed power prior borrowing approach with $\delta=0.5$. Figure \ref{fig:TwoSampleFix} (a) shows $\alpha_{\text{B}}(\theta_t=\theta_c; d_E)=\Exp_{\theta_t=\theta_c}[\varphi_{\text{B}}(D;d_E)]$ as a function of the difference between the current control mean $\theta_c$ and the observed external control average $\bar{d_E}$. 
Due to the assumption that data from the current treatment and control arms, as well as the external control data all have the same (known) standard deviation, this difference can be standardized by $\sigma$ to provide a more general figure.

Figure \ref{fig:TwoSampleFix}(a) shows that when $\theta_c$ is far smaller than $\bar{d_E}$, i.e. $\bar{d_E}$ is far in the alternative, $\frac{\theta_c-\bar{d_E}}{\sigma}$ is strongly negative and the probability of falsely rejecting $H_0$ is small, i.e., $\Exp_{\theta_t=\theta_c}[\varphi_{\text{B}}(D;d_E)]$ is close to $0$ because the current treatment is unlikely better than the external control. The more $\bar{d_E}$ moves towards small values, the higher the probability of falsely rejecting $H_0$, with $\Exp_{\theta_t=\theta_c}[\varphi_{\text{B}}(D;d_E)]$ going to $1$ for large $\frac{\bar{\theta_c-d_E}}{\sigma}$ because the external control increases the difference between treatment and control arm even if the current control and treatment data are similar. 

On the line $\theta_t - \theta_c= \theta_1=1$, power with borrowing is monotonically increasing as well.  In case $\bar{d_E}$ and $\theta_c$ are identical, $\alpha_{\text{B}}(\theta_c;d_E)$ is smaller than the nominal level of the test without borrowing, $\alpha=0.025$, because more control observations go into the evaluation of the two-sample test. Simultaneously, power at this point is higher than power without borrowing due to the same reason, a higher effective sample size in the control. Hence a gain in both frequentist operating characteristics, T1E rate and power, is observed at this point. The region in which gains are possible in both directions was termed the 'sweet spot' by Viele et al. \cite{Viele2014}.  

According to its definition, frequentist T1E rate is the maximum of the rejection probability for all parameters in $\Theta_0$, $\underset{\theta \in \Theta_0}{\max} \{\Exp_{\theta}[\varphi_{\text{B}}(D;d_E)]\}$ \cite{Lehmann1986}. In this situation the maximum is $\alpha_{\text{B}}(d_E)=1$. Hence the power of $\varphi_{\text{B}}$ cannot exceed that of the test calibrated to borrowing. Following the recommended Algorithm 1, Figure \ref{fig:TwoSampleFix}(b) shows again $\alpha_{\text{B}}(\theta_c=\theta_t;d_E)$ and the power difference $\Exp_{\theta_1}[\varphi_{\text{B}}(D;d_E)] - \Exp_{\theta_1}[\varphi^{(\alpha_{\text{B}}(d_E))}(D)]$ on the line $\theta_t-\theta_c=1$. The power difference is negative for all constellations of $d_E$ and $\theta_c$.

\subsubsection{Empirical Bayes power prior approach}
\begin{figure}[ht!]
\begin{center}

\includegraphics[scale=0.7]{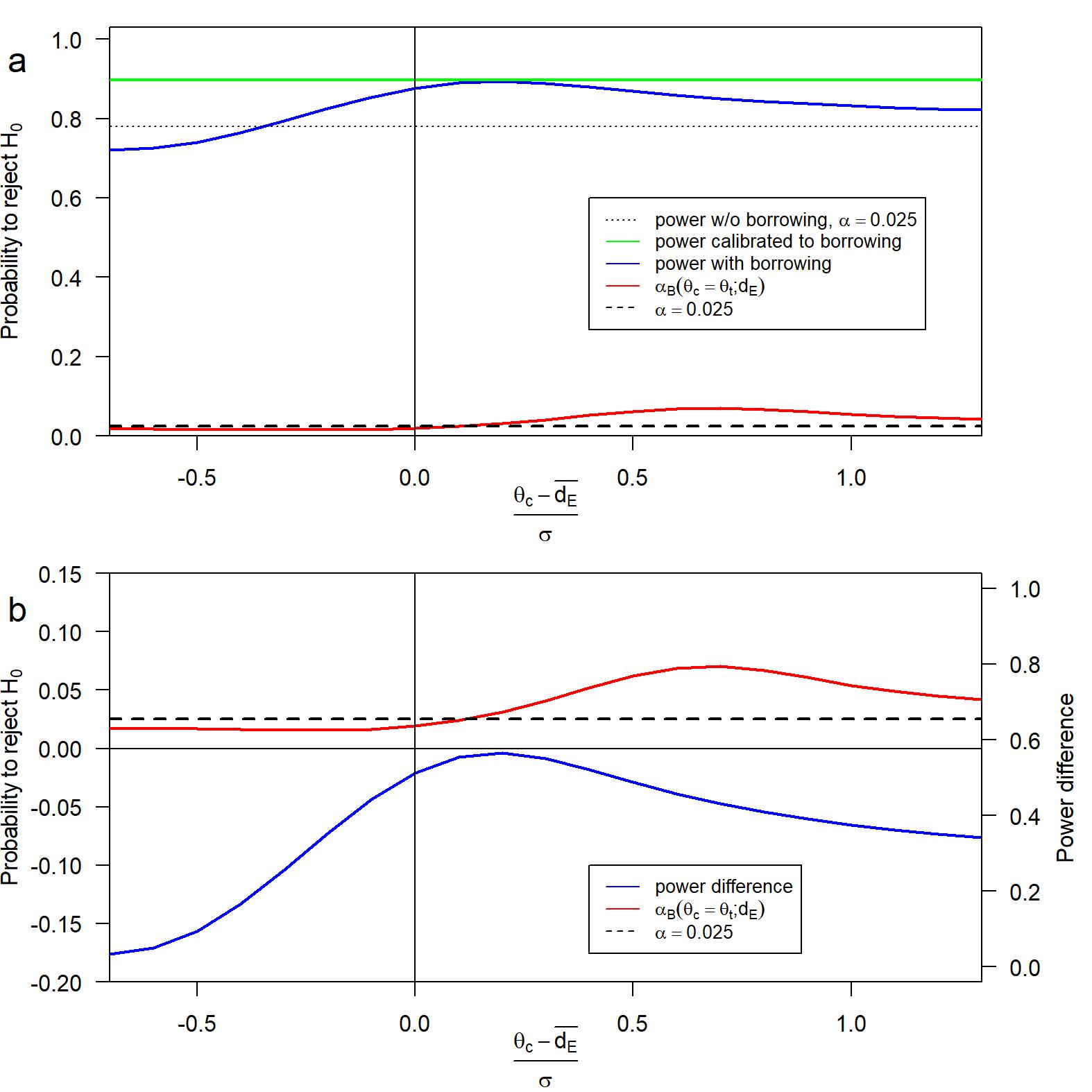}
\caption{Two-arm trial with Empirical Bayes power prior borrowing of external to current control. (a) $H_0$ rejection probabilities in the null hypothesis $\theta_c=\theta_t$, $\alpha_{\text{B}}(\theta_c=\theta_t;d_E)$, and on the alternative $\theta_t-\theta_c=1$ for $\varphi_{\text{B}}$ and $\varphi^{(\alpha_{\text{B}}(d_E))}$; (b)  $\alpha_{\text{B}}(\theta_c=\theta_t;d_E)$ and power difference $\Exp_{\theta_t-\theta_c=1}[\varphi_{\text{B}}(D;d_E)]- \Exp_{\theta_t-\theta_c=1}[\varphi^{(\alpha_{\text{B}}(d_E))}(D)]$.  \label{fig:TwoSampleEB}}
\end{center}
\end{figure}

When using an empirical Bayes power prior approach to augment the current with external control data, external data are discarded in case of prior-data conflict. Figure \ref{fig:TwoSampleEB} (a) again shows $\alpha_{\text{B}}(\theta_c=\theta_t;d_E)$ as well as the power with borrowing and the power of the test calibrated to borrowing, i.e. the power of the test with level $\alpha_B(d_E)= \underset{\theta_c=\theta_t }{\max} \{\alpha_{\text{B}}(d_E)\}$. Maximum T1E rate is attained at about $\frac{\theta_c-\bar{d_E}}{\sigma} =0.7$ with a value of $0.07$.  

Compared to the fixed power prior approach used in Section (\ref{Hybrid fixed PP}), the ability of Empirical Bayes power prior approach to adapt to prior-data conflict becomes apparent as type I deflation is less strong for values of $\theta_c$ much smaller than $\bar{d_E}$  and inflation is far less pronounced for $\theta_c$ much larger than $\bar{d_E}$. The power of the test calibrated to borrowing, $\varphi^{(\alpha_{\text{B}}(d_E))}$, however, is always larger than the power of $\varphi_{\text{B}}$, hence the power difference shown in  Figure \ref{fig:TwoSampleEB} (b) is negative. 

Note that in contrast to the one-sample situation where the borrowing method given a fixed external data set $d_E$ and a specific alternative $\theta$ is characterized by a point $(\alpha_{\text{B}}(d_E), \Exp_{\theta_1}[\varphi_{\text{B}}(D;d_E)] - \Exp_{\theta_1}[\varphi^{(\alpha_{\text{B}}(d_E))}(D)])$, in the two-arm hybrid control situation it is characterized by two lines $\alpha_{\text{B}}(\theta_c=\theta_t;d_E)$ and  $\Exp_{\theta_t-\theta_c=1}[\varphi_{\text{B}}(D;d_E)]- \Exp_{\theta_t-\theta_c=1}[\varphi^{(\alpha_{\text{B}}(d_E))}(D)]$. Since $\theta_c$ is unknown, T1E rate has to be taken as the maximum on the border of $\Theta_0$, $\alpha_{\text{B}}(d_E) = \underset{\theta_c=\theta_t }{\max} \{\alpha_{\text{B}}(d_E)\}$.

\section{Borrowing from random external data}\label{random external}
There may be circumstances in which frequentist operating characteristics of a borrowing method are of interest when not only the current data are random observations but also the external data should be considered as random. This situation may occur if, e.g., external data come from a real world data set and patients from this data set are selected for borrowing on the basis of similarity to current patient characteristics as reflected by covariable information (e.g., \cite{Wang2019}). In this case, the trial endpoint information from external patients can be considered as random in the same way as trial endpoint information from patients of the current trial. 
To investigate the performance of a borrowing method in this situation, the data generating parameter, $\theta_E$, for the external information, however, should still be considered as independent of the parameter of the current trial $\theta$. The procedure in Algorithm \ref{algo:random} is recommended to investigate and report the frequentist operating characteristics of a borrowing method in such a context.

\begin{algorithm}
\caption{Procedure for random external data}
\begin{algorithmic}
\State(R1) Select $\theta_E$ as parameter for external data.
\State(R2) Select $\theta \in \Theta_0$ for T1E rate calculation, typically $\theta=\theta_0$ on the border of $\Theta_0$. 
\State(R3) Select $\theta_1 \notin \Theta_0$ for power calculation.
\State(R4) Repeat for a sufficient number of times $n_{\text{sim}}$:
\begin{itemize} 
\item[] (R4.a) Generate one external data set $d_E$ from $D_E \sim f_{\theta_E} $.
 \item[] (R4.b) Generate one current data set $d_{\theta}$ from $D \sim f_{\theta} $, $\theta \in \Theta_0$ selected in (R2).
\item[] (R4.c) Record test decision $\varphi_{\text{B}}(d_{\theta};d_E)$.
\item[] (R4.d) Generate one current data sets $d_{\theta_1}$ from $D \sim f_{\theta_1}$.
\item[] (R4.e) Record test decision $\varphi_{\text{B}}(d_{\theta_1};d_E)$.
\end{itemize} 
\State(R5) Average test decisions from (R4.c) to obtain $\Exp_{\theta;\theta_E}[\varphi_{\text{B}}(D;D_E)], \theta \in \Theta_0$.
\State(R6) Average test decisions from (R4.e) to obtain $\Exp_{\theta_1;\theta_E}[\varphi_{\text{B}}(D;D_E)], \theta_1 \notin \Theta_0 $ for power evaluation.
\end{algorithmic}
\label{algo:random}
\end{algorithm}

If the rejection probability on the null hypothesis has to be evaluated not only at one point (e.g., in $\theta_0$ for the one-arm situation) but, e.g., on a line (as in case of the two-arm hybrid control trial in Section \ref{hybrid control trial}), multiple parameter values have to be selected in step (R2) to evaluate the maximum of the averaged test decisions in (R4) that we will denote by $\alpha_{\text{B}}(\theta_E)$.

On the same line of arguments as discussed in Section \ref{fixed external}, it is hard to compare the frequentist performance of the test without borrowing and the test with borrowing if they have different T1E rates. Again, for a fair comparison, the level of the test without borrowing is calibrated to  $\alpha_{\text{B}}(\theta_E)$, i.e., the difference $\Exp_{\theta_1;\theta_E}[\varphi_{\text{B}}(D;D_E)]- \Exp_{\theta_1}[\varphi^{(\alpha_{\text{B}}(\theta_E))}(D)]$ is evaluated.

\subsection{One-sided one-arm normal test}
As in Section \ref{one-sided one-arm}, we will first consider the situation of a one-arm trial with normally distributed endpoint to test $H_0: \theta \leq \theta_0 = 0 \text{ vs. } H_1: \theta > 0$. Again assume current data $D_i \sim N(\theta, 1), i=1,...,n$, and external data $D_{E,j} \sim N(\theta_E, 1), j=1,...,n_E$, with sample sizes $n=25$ and $n_E=20$. 
Note that in this situation the quantity $\alpha_{\text{B}}(\theta_E)$ is the average $\alpha_{\text{B}}(d_E)$ obtained in step (F3.b) of Algorithm 1, i.e. $\alpha_{\text{B}}(\theta_E)=\Exp_{\theta_E}[\alpha_{\text{B}}(d_E)]$.
Algorithm 2 results in two values for each parameter $\theta_E$ of the external data, $\alpha_{\text{B}}(\theta_E)$ and $\Exp_{\theta_1;\theta_E}[\varphi_{\text{B}}(D;D_E)]- \Exp_{\theta_1}[\varphi^{(\alpha_{\text{B}}(\theta_E))}(D)]$ (for $\theta_1 \notin \Theta_0$ selected for power evaluation).

When borrowing with a fixed power prior with $\delta=0.5$,  $n_{\text{sim}}=100,000$ simulations with $\theta_E=0$ resulted in
$\alpha_{\text{B}}(\theta_E=0) = 0.017$ and $\Exp_{\theta_1=0.5;\theta_E=0}[\varphi_{\text{B}}(D;D_E)] =  0.566$ compared to $\Exp_{\theta_1=0.5}[\varphi^{(\alpha_{\text{B}}(0.0))}(D)]=0.649$, i.e. a power difference of $-0.082$.
For $\theta_E=0.5$ the values were $\alpha_{\text{B}}(\theta_E=0.5) = 0.114$ and $\Exp_{\theta_1=0.5;\theta_E=0.5}[\varphi_{\text{B}}(D;D_E)] =  0.860$ compared to $\Exp_{\theta_1=0.5}[\varphi^{(\alpha_{\text{B}}(0.5))}(D)]=0.848$, i.e. a power difference of $-0.030$.
In this simple case of one-sided one-sample normal test with fixed power prior approach, the observations made above that the test calibrated to borrowing has higher power than the test with borrowing can be shown analytically, see Appendix. 

The situation when borrowing using an Empirical Bayes power prior approach is very similar.  With $n_{\text{sim}}=100,000$ simulations, using $\theta_E=0$ resulted in $\alpha_{\text{B}}(\theta_E=0) =  0.030$ and $\Exp_{\theta_1=0.5;\theta_E=0.5}[\varphi_{\text{B}}(D;D_E)] =  0.676$ compared to $\Exp_{\theta_1=0.5}[\varphi^{(\alpha_{\text{B}}(0.0))}(D)]=0.730$, i.e. a power difference of $-0.054$.
For $\theta_E=0.5$ the values were $\alpha_{\text{B}}(\theta_E=0.5) = 0.113$ and $\Exp_{\theta_1=0.5;\theta_E=0.5}[\varphi_{\text{B}}(D;D_E)] =  0.875$ compared to $\Exp_{\theta_1=0.5}[\varphi^{(\alpha_{\text{B}}(0.5))}(D)]=0.901$, i.e. a power difference of $-0.025$. For these selected cases we observe again that the test calibrated to borrowing has higher power than the test with borrowing. 

\subsection{One-sided two-arm normal test with borrowing to control arm}
\begin{figure}[ht!]
\begin{center}

\includegraphics[scale=0.7]{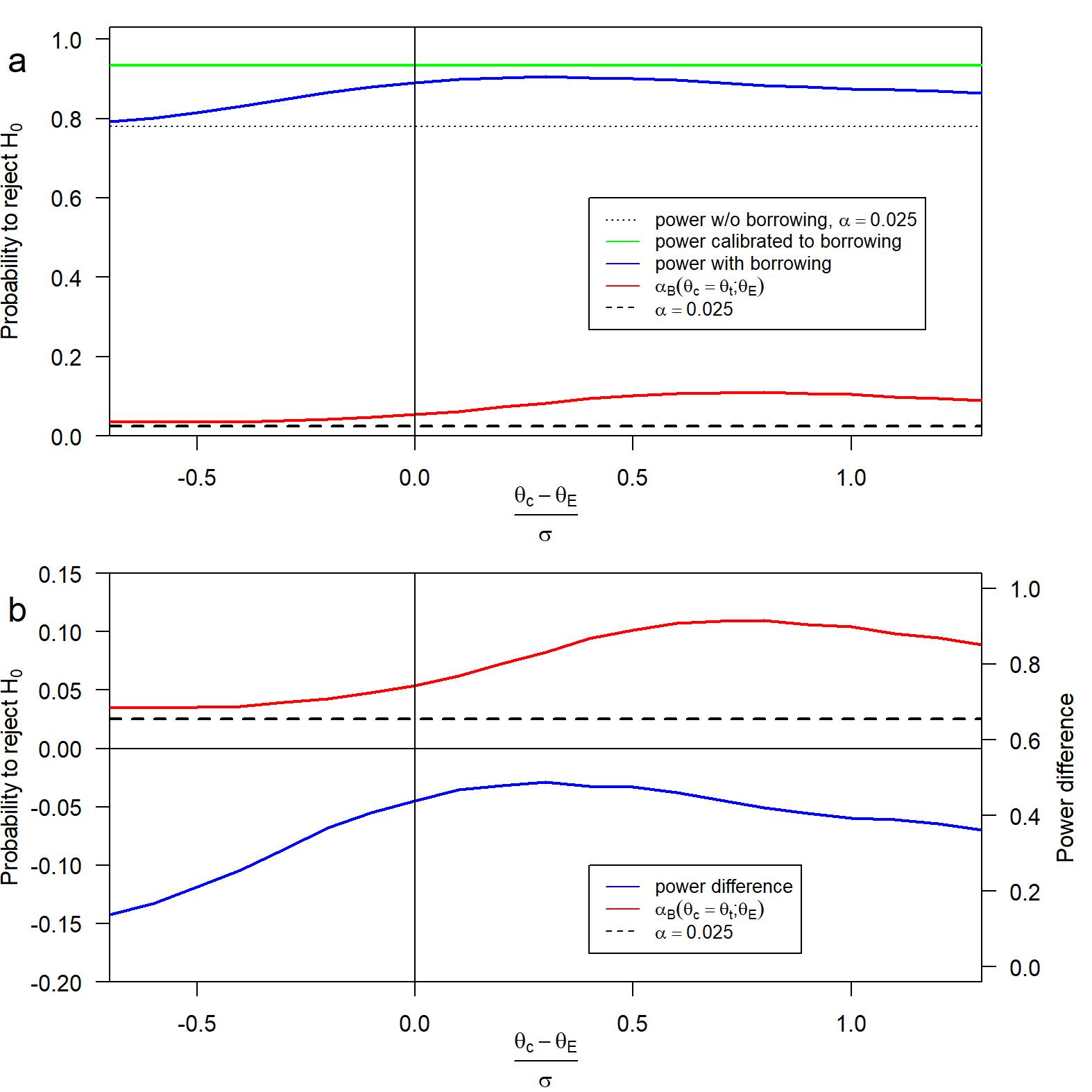}
\caption{Two-arm trial with Empirical Bayes power prior borrowing of external control to current control, regarding external data as random. (a) $H_0$ rejection probabilities in the null hypothesis $\theta_c=\theta_t$, $\alpha_{\text{B}}(\theta_c=\theta_t;\theta_E)$, and on the alternative $\theta_t-\theta_c=1$ for $\varphi_{\text{B}}$ and $\varphi^{(\alpha_{\text{B}}(\theta_E))}$; (b) $\alpha_{\text{B}}(\theta_c=\theta_t;\theta_E)$ and power difference $\Exp_{\theta_t-\theta_c=1}[\varphi_{\text{B}}(D;\theta_E)]- \Exp_{\theta_t-\theta_c=1}[\varphi^{(\alpha_{\text{B}}(\theta_E))}(D)]$.  \label{fig:TwoSampleEBRandom}}
\end{center}
\end{figure}

For the situation described in Section (\ref{hybrid control trial}), testing $H_0: \theta_t - \theta_c \leq 0 \text{ vs. } H_1: \theta_t -\theta_c > 0$, and again using an Empirical Bayes power prior approach for borrowing to control, the resulting plots for rejection probabilities are shown in Figure \ref{fig:TwoSampleEBRandom}. The overall shape is very similar to Figure \ref{fig:TwoSampleEB}. Comparing Figure \ref{fig:TwoSampleEB}(b) and Figure \ref{fig:TwoSampleEBRandom}(b), both the T1E rate and the power difference curves are slightly attenuated to $0$ when external data is considered random. 

\section{Discussion}\label{Discussion}
We propose a procedure to investigate and report the frequentist operating characteristics of methods for borrowing from external data to a current clinical trial. It is well-known that borrowing from external information may lead to the alteration of T1E rate. Since the relationship between T1E rate and power of a test is non-linear and not intuitive, we propose to evaluate T1E rate of the test with borrowing from external data, and to then calibrate the test without borrowing to this T1E rate to allow for a fair comparison of the power of the test with and without borrowing. It may seem counterintuitive to calibrate the test without borrowing. The reason for this is that situations may occur, in which the test with borrowing cannot be modified to control the original T1E rate, e.g. as shown in the hybrid control two-arm trial with fixed power prior borrowing where T1E rate is $1$.  On the other hand, tests without borrowing are available at any significance level. The proposed approach is illustrated in two clinical trial situations with normal endpoint, a one-arm and a hybrid control two-arm trial. The exemplary borrowing method used for illustration is a fixed and an Empirical Bayes power prior approach. Note, however, that the proposed procedure does not depend on the actual borrowing method that is used.

The assessment of the frequentist operating characteristics in the hybrid control two-arm situation is less straightforward than in the one-arm situation. According to its definition, the frequentist T1E rate is the maximum $H_0$ rejection probability over all parameter constellations in the null hypothesis. Hence, when borrowing, all parameter constellations of the current control and treatment arm as well as of the external data have to be considered to determine the T1E rate that is used for calibrating the test without borrowing. In the normal endpoint situation, considering all constellations is easily carried out using analytical calculations. With other endpoints (e.g., time-to-event) it is, however, necessary to compute $H_0$ rejection probabilities in all null hypothesis constellations, and subsequently compare the power with borrowing to the power of the test calibrated to borrowing.  

Calculation of power and T1E rate is usually associated with inaccuracy due to numerical integration or Monte Carlo approaches. Numerical inaccuracy can be propagated when calculating the power of a test calibrated to T1E rate of the test with borrowing. For a valid comparison of power between different approaches it is hence crucial to restrict numerical inaccuracy due to integration as much as possible.

Our approach can be used for the situation in which external data is fixed, as it is usually the case when a trial is designed with the plan to borrow from historical data. In this case, $\Exp_{\theta}[\varphi_{\text{B}}(D;D_E)\given D_E=d_E]$ is evaluated. A variant of the approach can also be used when external data should be considered as random, as it is the case when borrowing from data that were obtained from a real world data set by selecting patients without considering their outcome but evaluating similarity to the current patients. This is the situation in which $\Exp_{\theta; \theta_E}[\varphi_{\text{B}}(D;D_E)]$ is evaluated. For the special case of using a power prior approach with fixed $\delta$ parameter, we showed analytically that instead of a power gain, power is indeed decreased when borrowing. Using the Empirical Bayes power prior approach, numerical examples point in the same direction.

Occasionally operating characteristics of borrowing approaches have been investigated under the assumption that external data are random (as considered in Section \ref{random external}) and that in addition $\theta$ and $\theta_E$ are jointly generated by a statistical model. Most likely the test of interest in this case is still $H_0: \theta \in \Theta_0 \text{ vs. } H_1: \theta \notin \Theta_0$, i.e., it addresses $\theta$ and not the external parameter $\theta_E$. Hence, in this context, definition of T1E rate and power of the test are not entirely clear. 
If, however, the hypotheses of interest also comprise the parameter generating the external data, the test without and the test with borrowing would address different hypotheses.
A situation in which a joint evaluation of different parameters is required is given by, e.g., basket trials, where the same treatment is investigated in different patients subgroups in parallel. 
In this situation, however, the set of hypotheses and operating characteristics that are investigated is different from those in the case of external borrowing to a current trial discussed here. 

The idea that for test comparison, power should only be compared for tests with the same size (i.e. same T1E rate), has been put forward by Lloyd \cite{Lloyd2005} earlier. He suggested to use receiver operating characteristic (ROC) curves showing power versus size to compare tests. For example in the one-sample case, one would select a $d_E$ and derive the ROC curve by varying the threshold $c_{d_E}$ to generate different T1E rates and record the corresponding power, resulting in a bundle of ROC curves for different values of $d_E$. For the examples shown in Fig. \ref{fig:OneSample_grid} (a) and (b), all ROC curves would coincide, whereas for the situation in Fig. \ref{fig:OneSample_grid} (c), the test with borrowing would result in ROC curves below the one for the test without borrowing for $0.06 \leq d_E \leq 0.14$ but coinciding with the ROC for the test without borrowing for all other $d_E$. However, this approach may require more computing time without providing additional information compared to our approach and visualization.  

A recent approach proposes the use of relative likelihood ratios for neutral comparison of statistical tests to enable accurate comparison of the trade-off between power and size between between competing tests\cite{huang2023}. Unless one test is clearly superior in terms of positive and negative likelihood ratio, the application of this approach relies on the specific context, i.e., whether controlling size or increasing power is more important. 
Several additional proposals have recently tried to move away from strict control of T1E rate, e.g., by giving rationales for targeting a weighted sum of T1E and T2E rates instead\citep{grieve2015,pericchi2016,walley2021,calderazzo2022}. This can be accomplished by assigning weights representing the relative seriousness of T1E versus T2E rates (or relative trust on the null and alternative hypothesis), evaluated at $\theta_0 \in \Theta_0$ and $\theta_1 \notin \Theta_0$  \citep{grieve2015,walley2021}. In the context of one-sided one-arm testing, the latter can be directly translated into a Bayesian integrated risk minimization problem with a two-point mass prior, giving equal probability to $\theta_0$ and $\theta_1$, and assigning costs to T1E and T2E equal to the above weights. Bayesian prior probabilities can be incorporated in such an assessment; these can be used to assign a prior probability to each (point) hypothesis or to average across the support of $\theta$ itself. In the latter case, the costs can be reformulated so that optimal Bayesian decisions would minimize a weighted sum of average (or expected) T1E and T2E rates \cite{pericchi2016, calderazzo2022}. 

It is important to note that for one-sided one-arm testing of location parameters in exponential families, and for a given sample size, each approach would essentially select a specific rejection probability function, so that all approaches are equivalent if T1E rate (or power) is identical.  The Bayesian approach justifies T1E rate inflation by trust in the prior distribution (if it favors the alternative hypothesis), and leads in turn to a specific weighted sum of errors. In two-arm situations, however, Bayesian and frequentist tests will tend to differ. Without further constraints and under fixed borrowing, the maximum T1E rate of the Bayesian test can reach 1. Data-adaptive approaches such as the Empirical Bayes power prior can cap such an inflation, but cannot lead to uniform power gains, as  discussed earlier. In practice, overall T1E rate inflation is again justified by some trust in external/historical information, and gains in T1E rate and/or power will therefore be local, see also the discussion by Grieve \cite{Grieve2016}.

In general, to evaluate operating characteristics in a frequentist sense, no prior assumptions about the similarity of external and current parameter shall be made. If one wants to restrict investigations to situations where the external (control) and the current (control) data generating process are similar, the operating characteristics would not be frequentist but Bayesian in the sense that one would rely on the consistency of prior and current data. While the latter is a reasonable assumption if prior information is deemed reliable, it should be clearly stated when adopted.

\vspace*{1pc}


\section*{Appendix }
Here we show that fixed power prior borrowing from random external data in the one-sided one-arm normal test situation leads to a power loss. 

Consider the situation of a one-arm trial with normally distributed endpoint to test $H_0: \theta \leq \theta_0  \text{ vs. } H_1: \theta > \theta_0$. Assume $n$ current observations $D_i \sim N(\theta, \sigma)$, $\sigma$ known, and assume a prior $\pi \sim N(\mu_{\pi}, \sigma_{\pi}^2)$. 
This prior may, e.g., be obtained using a power prior approach with fixed $\delta \in [0,1]$, i.e., with $n_E$ external observations $D_{E,j} \sim N(\theta_E, \sigma^2)$, this results in $\mu_{\pi}= \bar{d}_E$ and $\sigma_{\pi}^2=\frac{\sigma^2}{\delta n_E}$. For abbreviation, set $\sigma_n=\sigma / \sqrt{n}$ and $\sigma_{n_E}=\sigma / \sqrt{n_E}$.

The posterior is then normally distributed with variance $\sigma_p^{-2} = \frac{1}{\sigma_{\pi}^2} + \frac{1}{\sigma_n^2}$
and mean $\mu_p = \sigma_p^2 \left( \frac{\mu_{\pi}}{\sigma_{\pi}^2} + \frac{\bar{d}}{\sigma_n^2 }\right)$. According to (\ref{Eq:phiBBayes}), $H_0$ is rejected when 
\begin{equation}
\label{Eq:A1}
\dfrac{\mu_p - \theta_0}{\sigma_p} > c 
\end{equation}
where $c$ is chosen such that T1E rate of the test without borrowing (i.e. with $\sigma_{\pi}=\infty$) is controlled at $\alpha$, implying that $c=z_{1-\alpha}$, the $(1-\alpha)$-quantile of the standard normal distribution. 
Rearranging terms, (\ref{Eq:A1}) is equivalent to 
\begin{equation}
\label{Eq:A2}
\dfrac{\bar{d} - \theta_0}{\sigma_n} + \frac{(\mu_{\pi}-\theta_0)\sigma_n}{\sigma_{\pi}^2} > z_{1-\alpha}\sqrt{1+\frac{\sigma_n^2}{\sigma_{\pi}^2}},
\end{equation}
see also the Supplement to \cite{Psioda2018}.

Now consider external data that we borrow from as random. Then $\mu_{\pi}$ on the lhs of (\ref{Eq:A2}) is no longer a fixed value but a random variable, i.e., for $\mu_{\pi} = \bar{d}_E$, we have
$\mu_{\pi} \sim N(\theta_E, \sigma_{n_E}^2)$.  
Let $x$ denote the lhs of (\ref{Eq:A2}), which is then a realization from the sum of two independent and normally distributed variables, $N\left((\theta - \theta_0)/\sigma_n,1 \right)$ and 
$N\left((\theta_E - \theta_0)\sigma_n/\sigma_{\pi}^2,\delta \sigma_n^2/\sigma_{\pi}^2\right)$ (note that because $\sigma_{\pi}^2=\sigma_E^2/(\delta n_E), \sigma_n^2 \sigma_{n_E}^2/\sigma_{\pi}^4=\delta \sigma_n^2/\sigma_{\pi}^2$), 
i.e., $X \sim N(\mu_x,\sigma_x^2)$ with 
$\mu_x(\theta)=(\theta - \theta_0)/\sigma_n + ((\theta_E-\theta_0)\sigma_n)/\sigma_{\pi}^2)$ and 
$\sigma_x^2=1+(\delta \sigma_n^2/\sigma_{\pi}^2)$.

With borrowing, $H_0$ is rejected if $x > z_{1-\alpha} \sqrt{1+ \sigma_n^2/\sigma_{\pi}^2}$, i.e., the rejection probability is given by 
\begin{equation}
1-\Phi \left(z_{1-\alpha}\frac{1}{\sigma_x}\sqrt{1+\frac{\sigma_n^2}{\sigma_{\pi}^2}}- \frac{\mu_x(\theta)}{\sigma_x} \right), 
\label{powborr}
\end{equation}

with $\Phi$ denoting the cumulative distribution function of the standard normal distribution. T1E rate with borrowing is reached in $\theta=\theta_0$, i.e. $$\alpha_{\text{B}}(\theta_E)=1-\Phi \left( z_{1-\alpha}\frac{1}{\sigma_x}\sqrt{1+\frac{\sigma_n^2}{\sigma_{\pi}^2}} -\frac{(\theta_E-\theta_0)\sigma_n}{\sigma_{\pi}^2\sigma_x} \right),$$ using the notation from Section \ref{random external}.
Denote by $z_{1-\alpha_{\text{B}}(\theta_E)}$ the corresponding quantile.

The test calibrated to borrowing rejects if $(\bar{d} - \theta_0)/\sigma_n > z_{1-\alpha_{\text{B}}(\theta_E)}$, the probability of which is  
\begin{equation}
1-\Phi \left( z_{1-\alpha_{\text{B}}(\theta_E)}-\frac{\theta - \theta_0}{\sigma_n} \right).
\label{powcal}
\end{equation}

The power of the test calibrated to borrowing (\ref{powcal}) evaluated at $\theta > \theta_0$ can be shown to be larger than the power of the test with borrowing (\ref{powborr}), by analytical transformations and observing that $\sigma_x > 1$ for $\delta>0$. In the notation of the manuscript, the above derivation shows that 
\begin{equation}
\nonumber
1-\Phi \left( z_{1-\alpha_{\text{B}}(\theta_E)}-\frac{\theta - \theta_0}{\sigma_n} \right)=\Exp_{\theta}[\varphi^{(\alpha_{\text{B}}(\theta_E))}(D)] > 1-\Phi \left(z_{1-\alpha}\frac{1}{\sigma_x}\sqrt{1+\frac{\sigma_n^2}{\sigma_{\pi}^2}}- \frac{\mu_x(\theta)}{\sigma_x} \right) =\Exp_{\theta; \theta_E}[\varphi_{\text{B}}(D;D_E)].
\end{equation}

\section*{Data availability statement}
Data sharing is not applicable to this article as no new data were created or analyzed in this study.

\clearpage

\end{document}


\begin{center}
\huge{\textbf{Supplementary material}}\\

\vspace{1cm}
\Large
\textit{Simulating and reporting frequentist operating characteristics of clinical trials that borrow external information}\\
\vspace{1cm}
\centering Annette Kopp-Schneider, Manuel Wiesenfarth, Leonhard Held and Silvia Calderazzo 
\end{center}
\normalsize
\vspace{1cm}

\section*{Extreme borrowing can lead to a test that is not UMP and consequently to power loss}
To illustrate that dynamic borrowing can lead to a test that is not UMP, and hence that power loss can be the result of borrowing, we consider an extreme and artificial setting. 
In case of a one-arm trial with normally distributed endpoint with known variance $\sigma=1$ (i.e., current data $D_i \sim N(\theta,1), i=1,...,n$), we test $H_0: \theta \leq \theta_0 = 0 \text{ vs. } H_1: \theta > 0$ and evaluate power at $\theta_1=0.5$. Sample size for current data is $n=25$. 
We use the Empirical Bayes power prior approach to borrow from $n_E=1000$ external data $D_{E,j} \sim N(\theta_E, 1), j=1,...,n_E$. 

\begin{figure}[ht!]
\begin{center}
\includegraphics[scale=0.7]{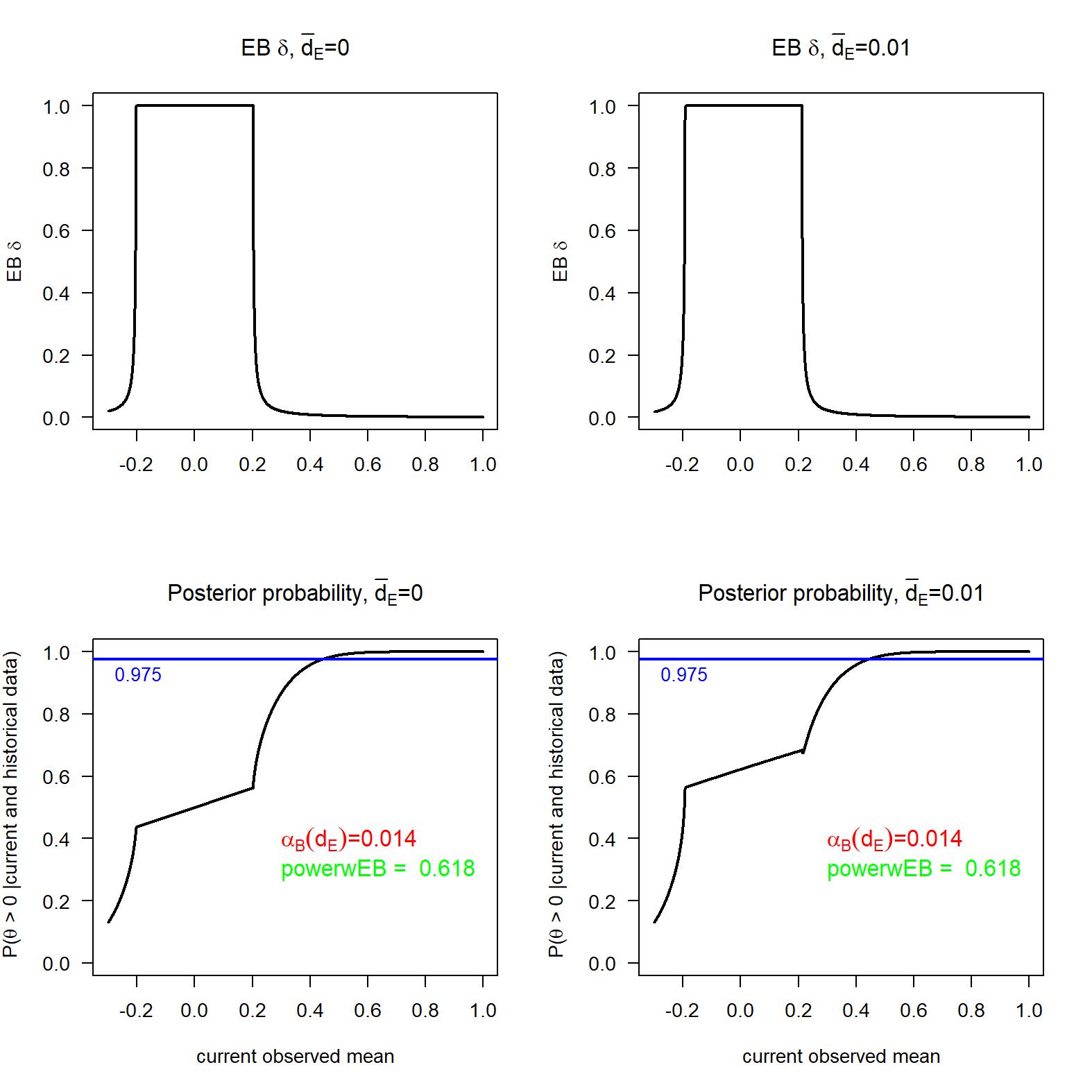}
\caption{$\hat{\delta}(d;d_E)$ (top) and the posterior probability $\Prob(\theta > \theta_0 \given  D=d; D_E = d_E)$ (bottom) for varying current data mean $\bar d$ for $\bar d_E=0$ (left) and  $\bar d_E=0.01$ (right). \label{fig:Extreme_EB0}}
\end{center}
\end{figure}

\begin{figure}[ht!]
\begin{center}
\includegraphics[scale=0.7]{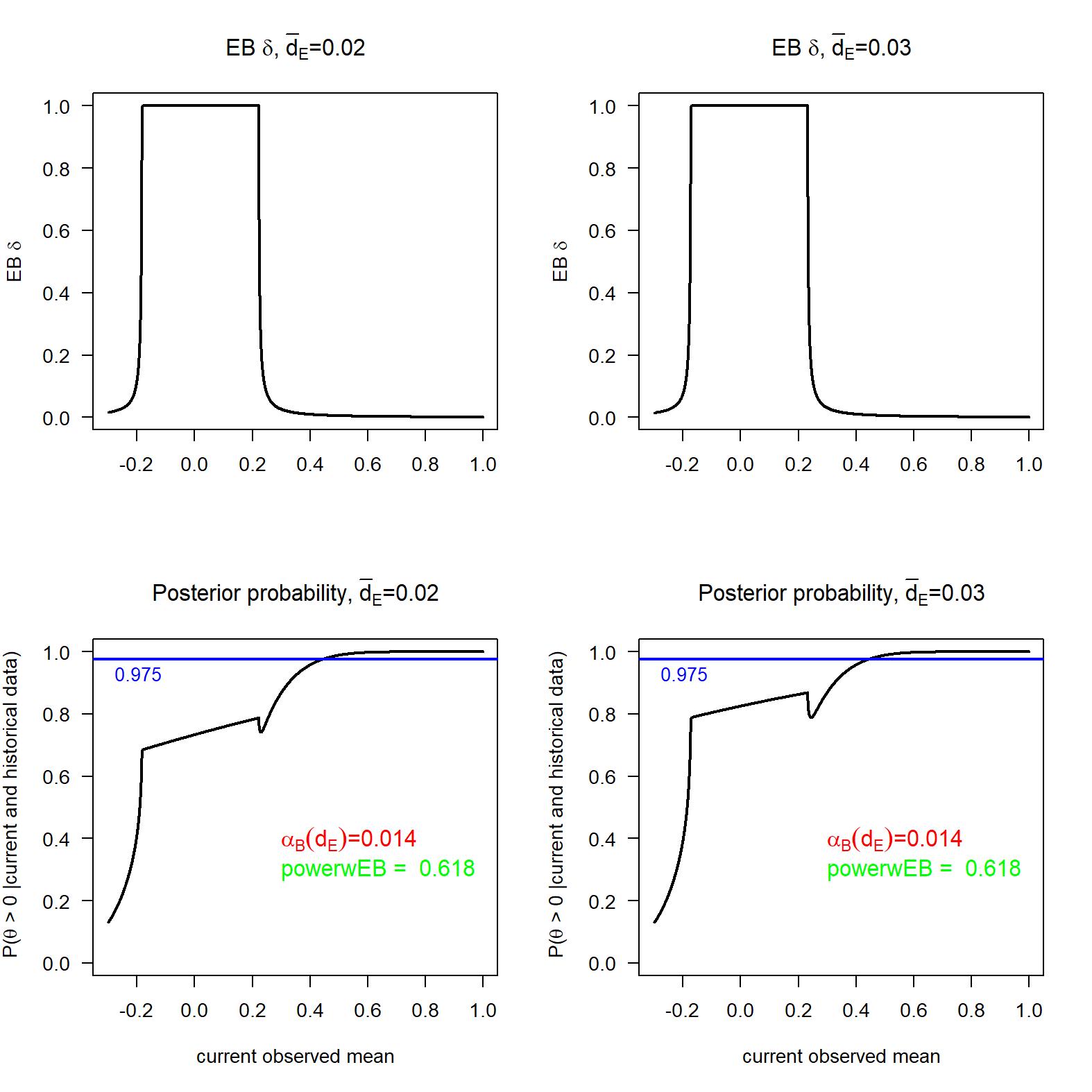}
\caption{$\hat{\delta}(d;d_E)$ (top) and the posterior probability $\Prob(\theta > \theta_0 \given  D=d; D_E = d_E)$ (bottom) for varying current data mean $\bar d$ for $\bar d_E=0.02$ (left) and  $\bar d_E=0.03$ (right). \label{fig:Extreme_EB1}}
\end{center}
\end{figure}

\begin{figure}[ht!]
\begin{center}
\includegraphics[scale=0.7]{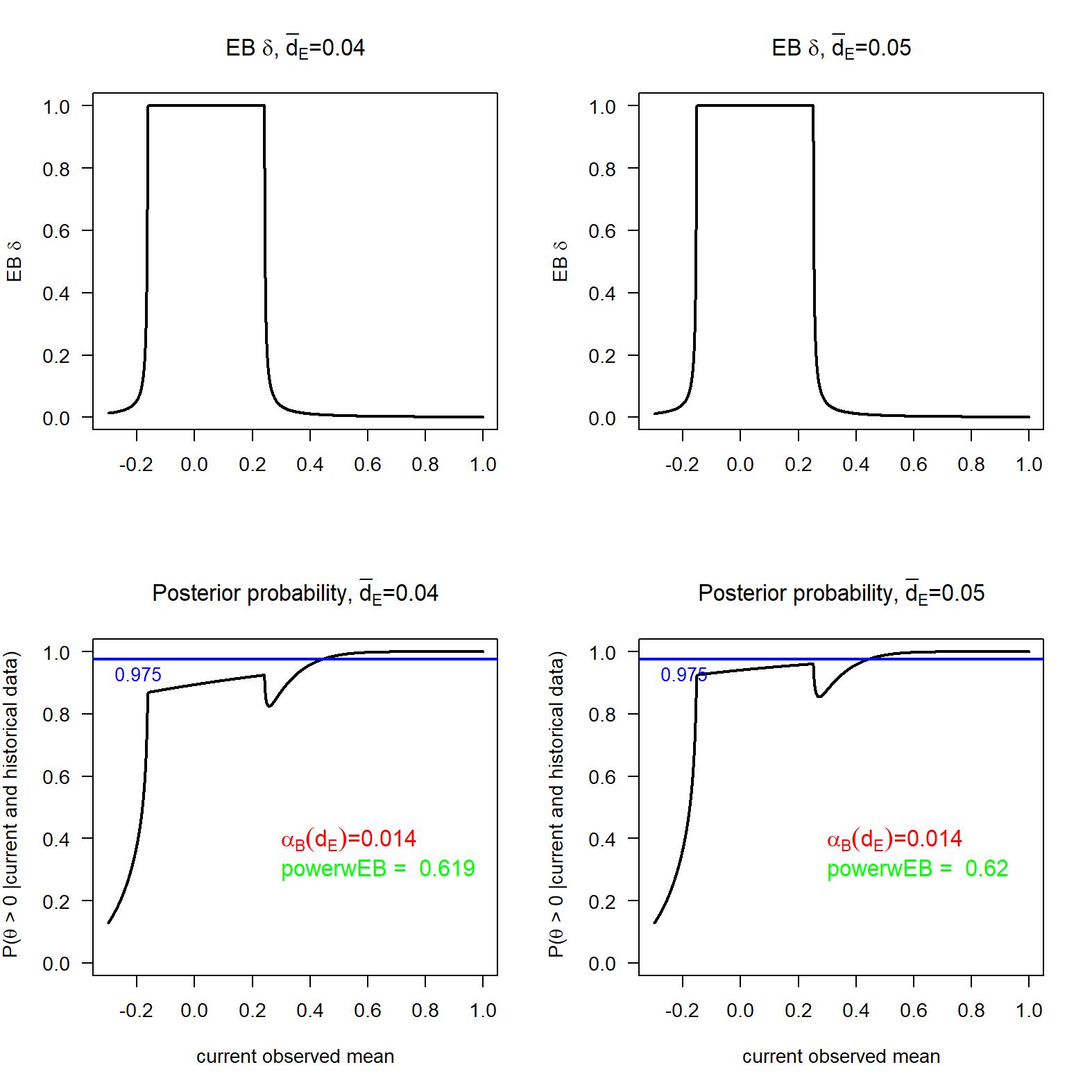}
\caption{$\hat{\delta}(d;d_E)$ (top) and the posterior probability $\Prob(\theta > \theta_0 \given  D=d; D_E = d_E)$ (bottom) for varying current data mean $\bar d$ for $\bar d_E=0.04$ (left) and  $\bar d_E=0.05$ (right). \label{fig:Extreme_EB2}}
\end{center}
\end{figure}

\begin{figure}[ht!]
\begin{center}
\includegraphics[scale=0.7]{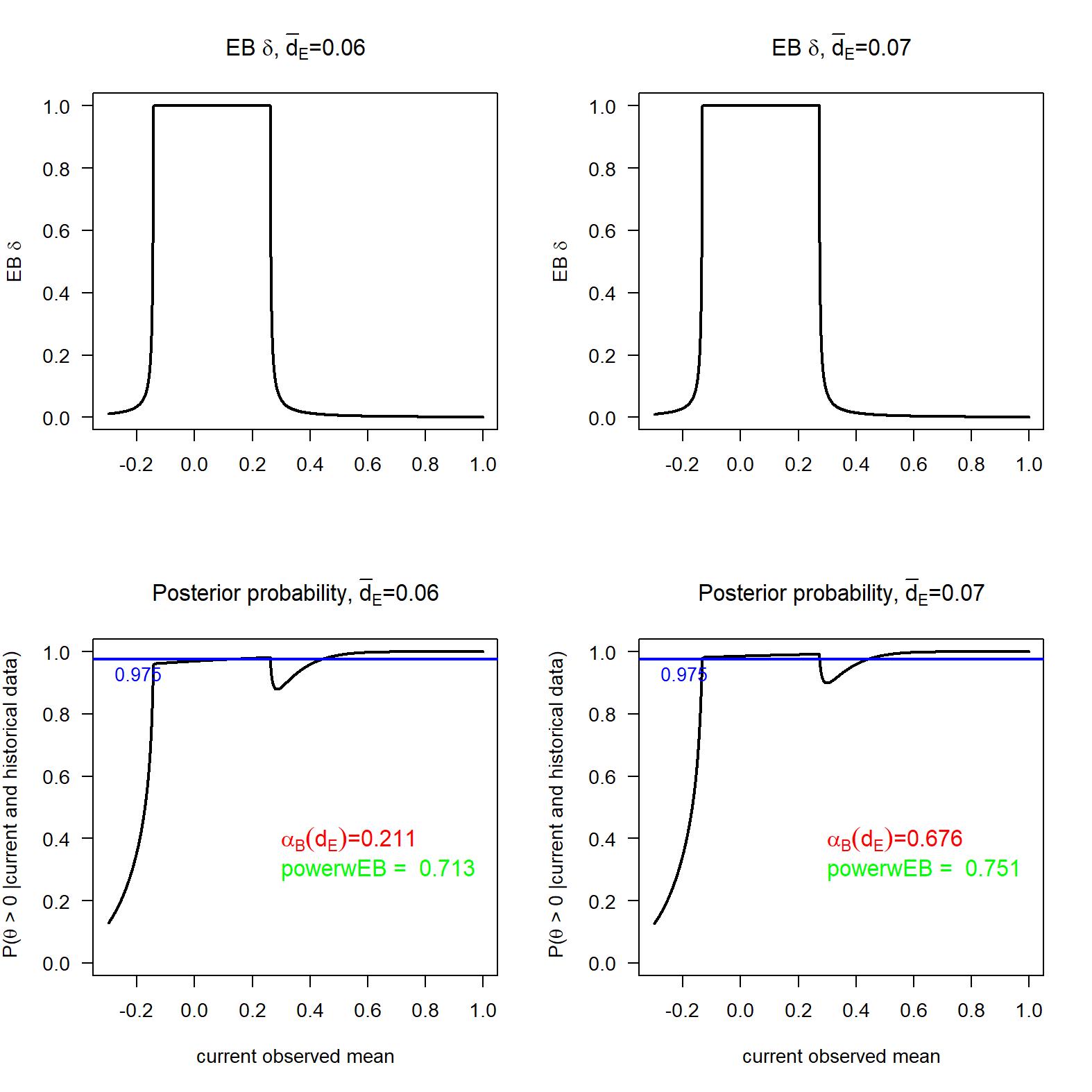}
\caption{$\hat{\delta}(d;d_E)$ (top) and the posterior probability $\Prob(\theta > \theta_0 \given  D=d; D_E = d_E)$ (bottom) for varying current data mean $\bar d$  for $\bar d_E=0.06$ (left) and  $\bar d_E=0.07$ (right). \label{fig:Extreme_EB3}}
\end{center}
\end{figure}

\begin{figure}[ht!]
\begin{center}
\includegraphics[scale=0.7]{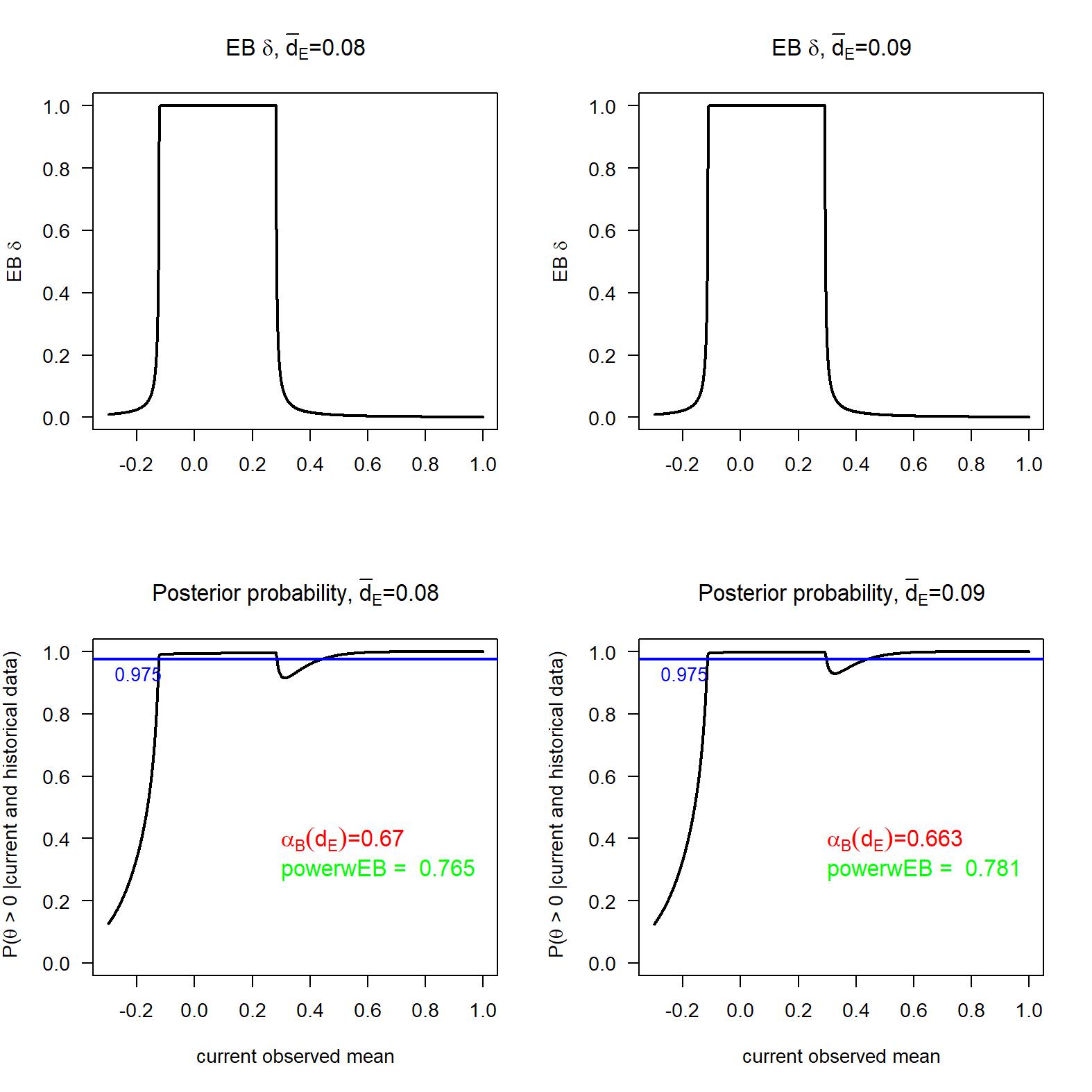}
\caption{$\hat{\delta}(d;d_E)$ (top) and the posterior probability $\Prob(\theta > \theta_0 \given  D=d; D_E = d_E)$ (bottom) for varying current data mean $\bar d$ for $\bar d_E=0.08$ (left) and  $\bar d_E=0.09$ (right). \label{fig:Extreme_EB4}}
\end{center}
\end{figure}

\begin{figure}[ht!]
\begin{center}
\includegraphics[scale=0.7]{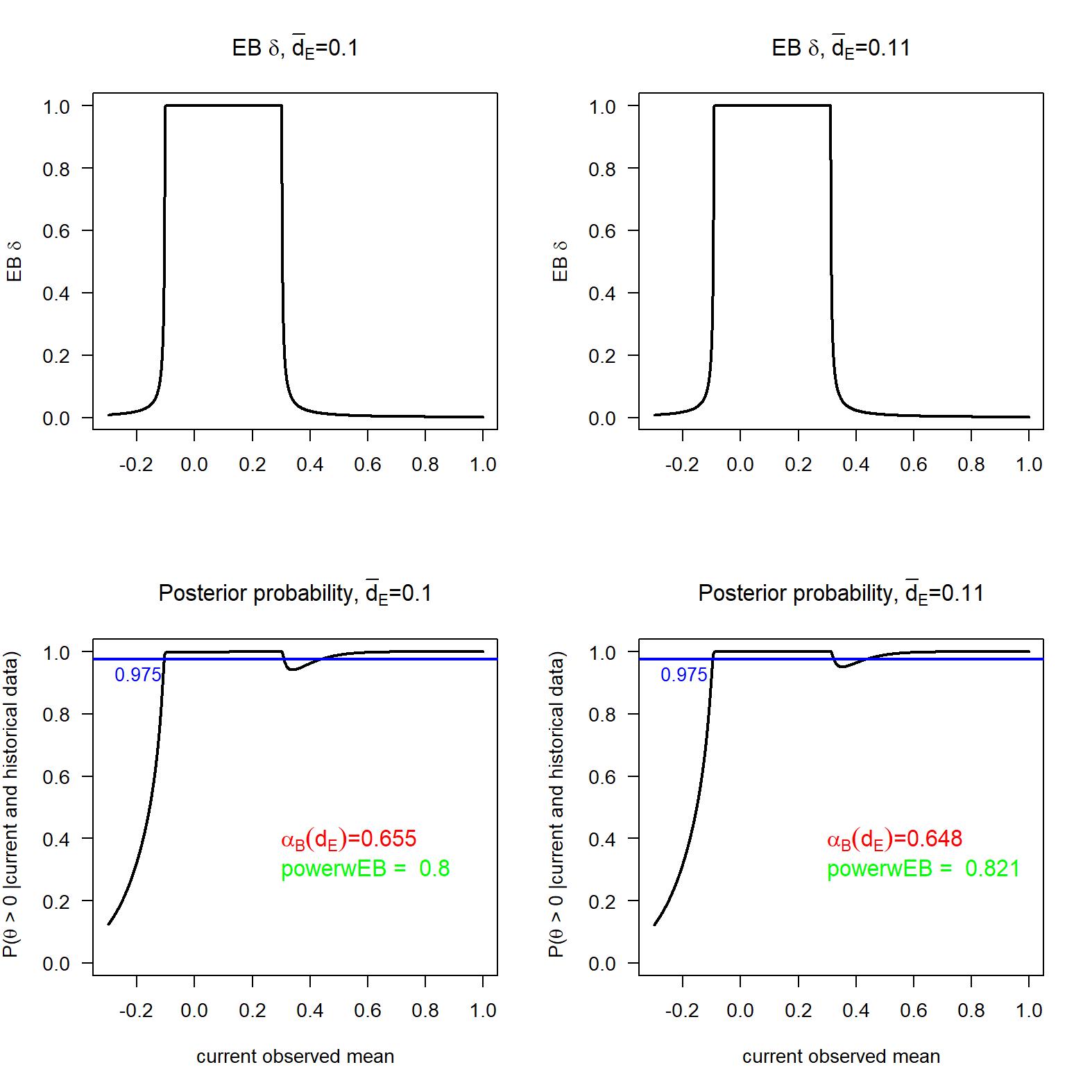}
\caption{$\hat{\delta}(d;d_E)$ (top) and the posterior probability $\Prob(\theta > \theta_0 \given  D=d; D_E = d_E)$ (bottom) for varying current data mean $\bar d$ for $\bar d_E=0.10$ (left) and  $\bar d_E=0.11$ (right). \label{fig:Extreme_EB5}}
\end{center}
\end{figure}

\begin{figure}[ht!]
\begin{center}
\includegraphics[scale=0.7]{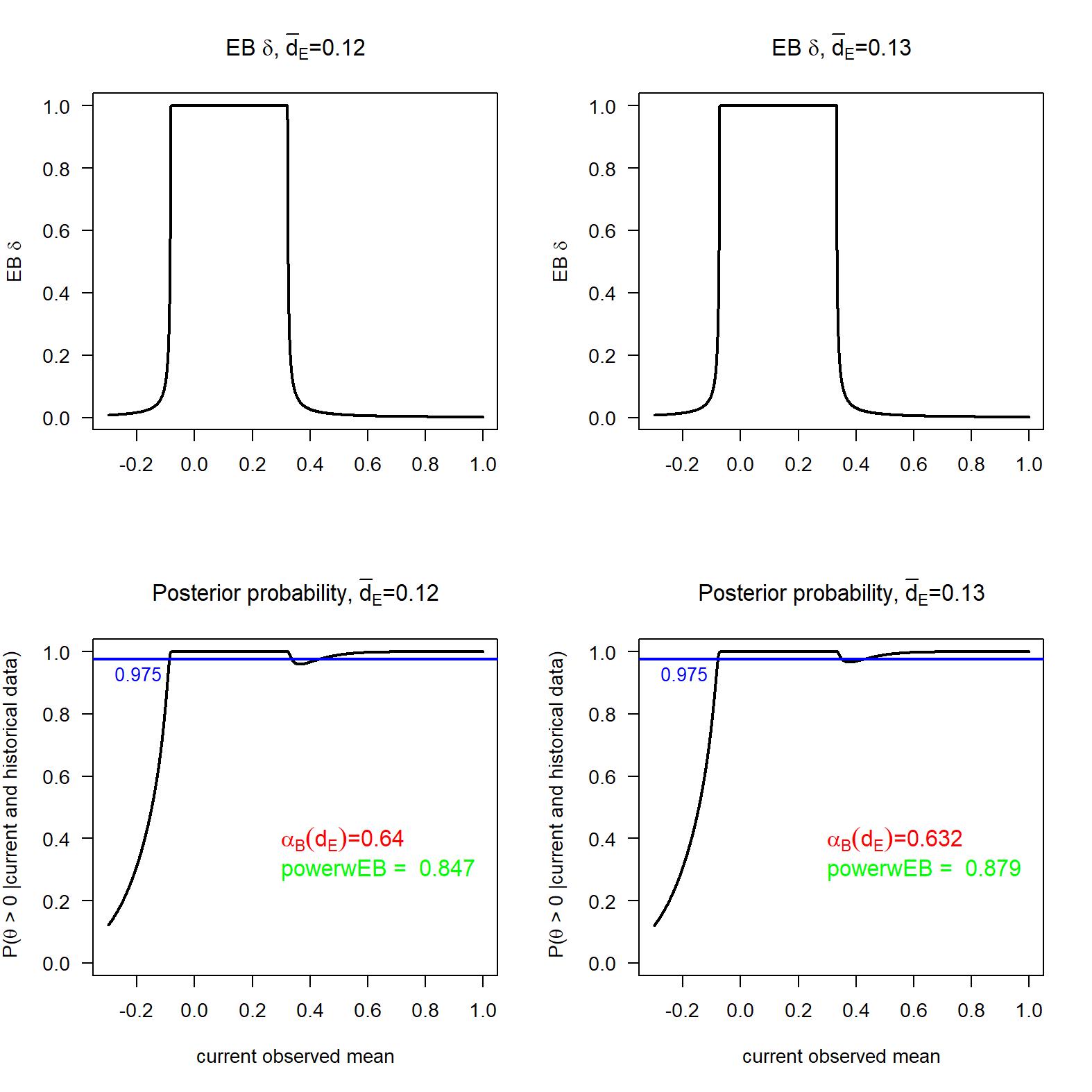}
\caption{$\hat{\delta}(d;d_E)$ (top) and the posterior probability $\Prob(\theta > \theta_0 \given  D=d; D_E = d_E)$ (bottom) for varying current data mean $\bar d$ for $\bar d_E=0.12$ (left) and  $\bar d_E=0.13$ (right). \label{fig:Extreme_EB6}}
\end{center}
\end{figure}

\begin{figure}[ht!]
\begin{center}
\includegraphics[scale=0.7]{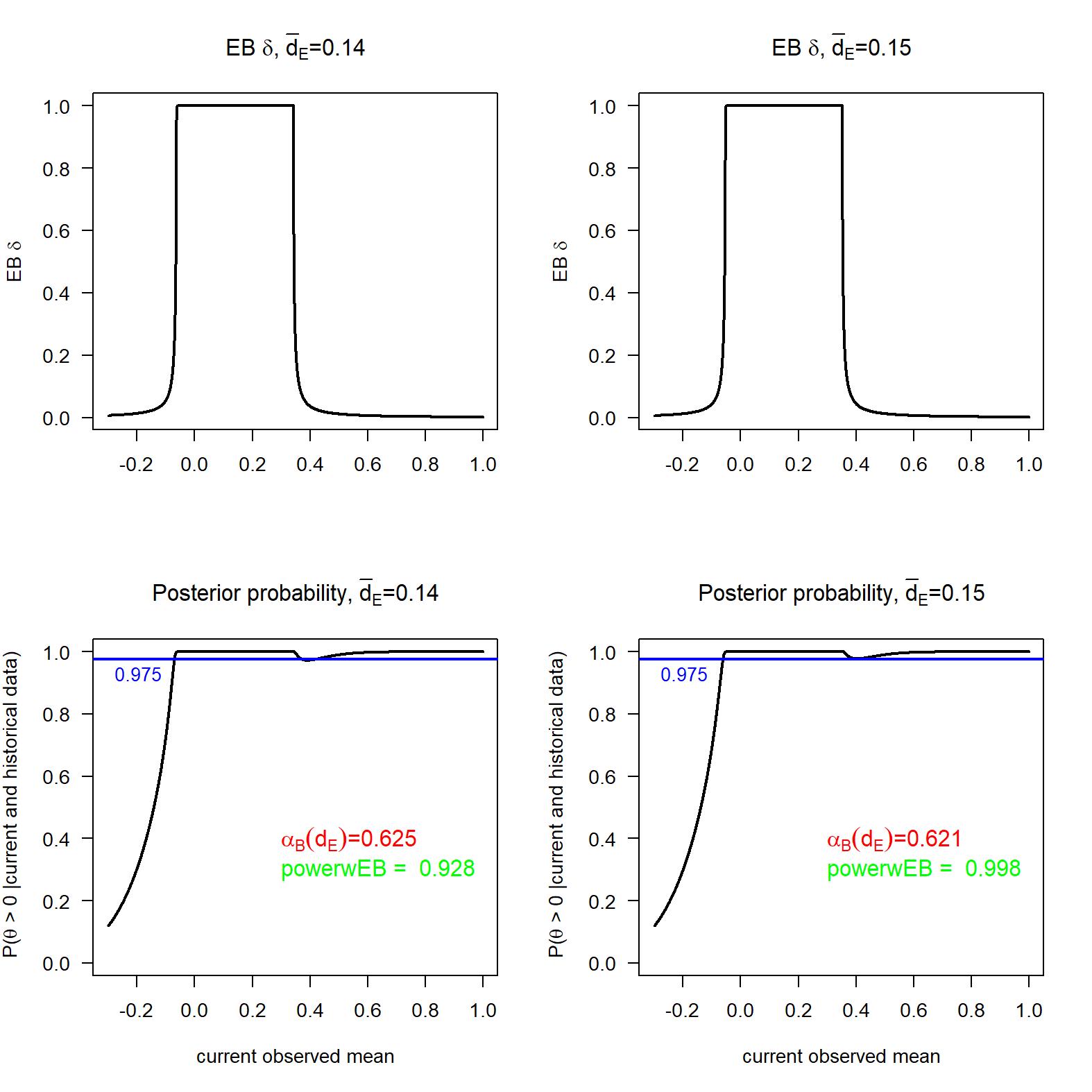}
\caption{$\hat{\delta}(d;d_E)$ (top) and the posterior probability $\Prob(\theta > \theta_0 \given  D=d; D_E = d_E)$ (bottom) for varying current data mean $\bar d$ for $\bar d_E=0.14$ (left) and  $\bar d_E=0.15$ (right). \label{fig:Extreme_EB7}}
\end{center}
\end{figure}

\begin{figure}[ht!]
\begin{center}
\includegraphics[scale=0.7]{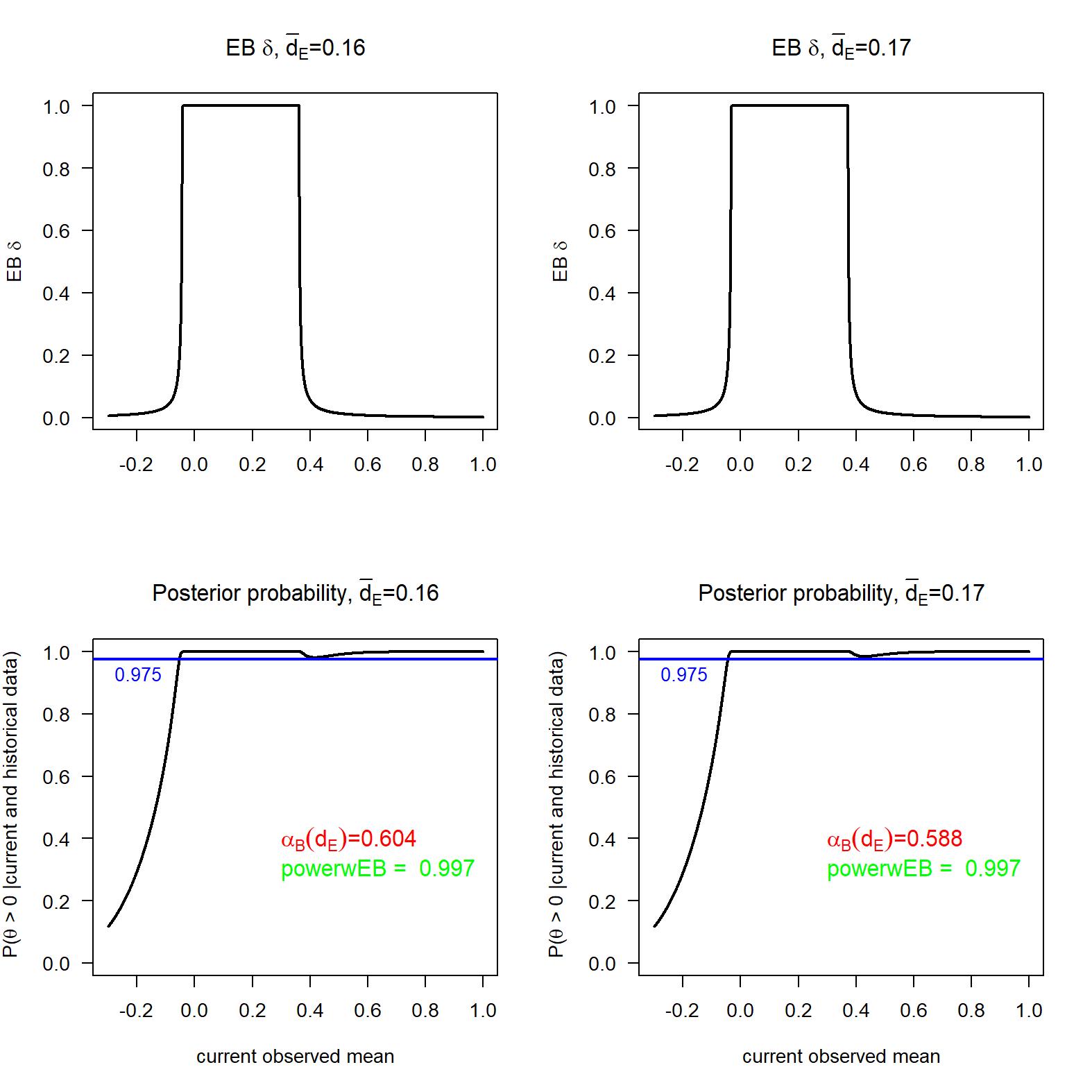}
\caption{$\hat{\delta}(d;d_E)$ (top) and the posterior probability $\Prob(\theta > \theta_0 \given  D=d; D_E = d_E)$ (bottom) for varying current data mean $\bar d$ for $\bar d_E=0.16$ (left) and  $\bar d_E=0.17$ (right). \label{fig:Extreme_EB8}}
\end{center}
\end{figure}

\begin{figure}[ht!]
\begin{center}
\includegraphics[scale=0.7]{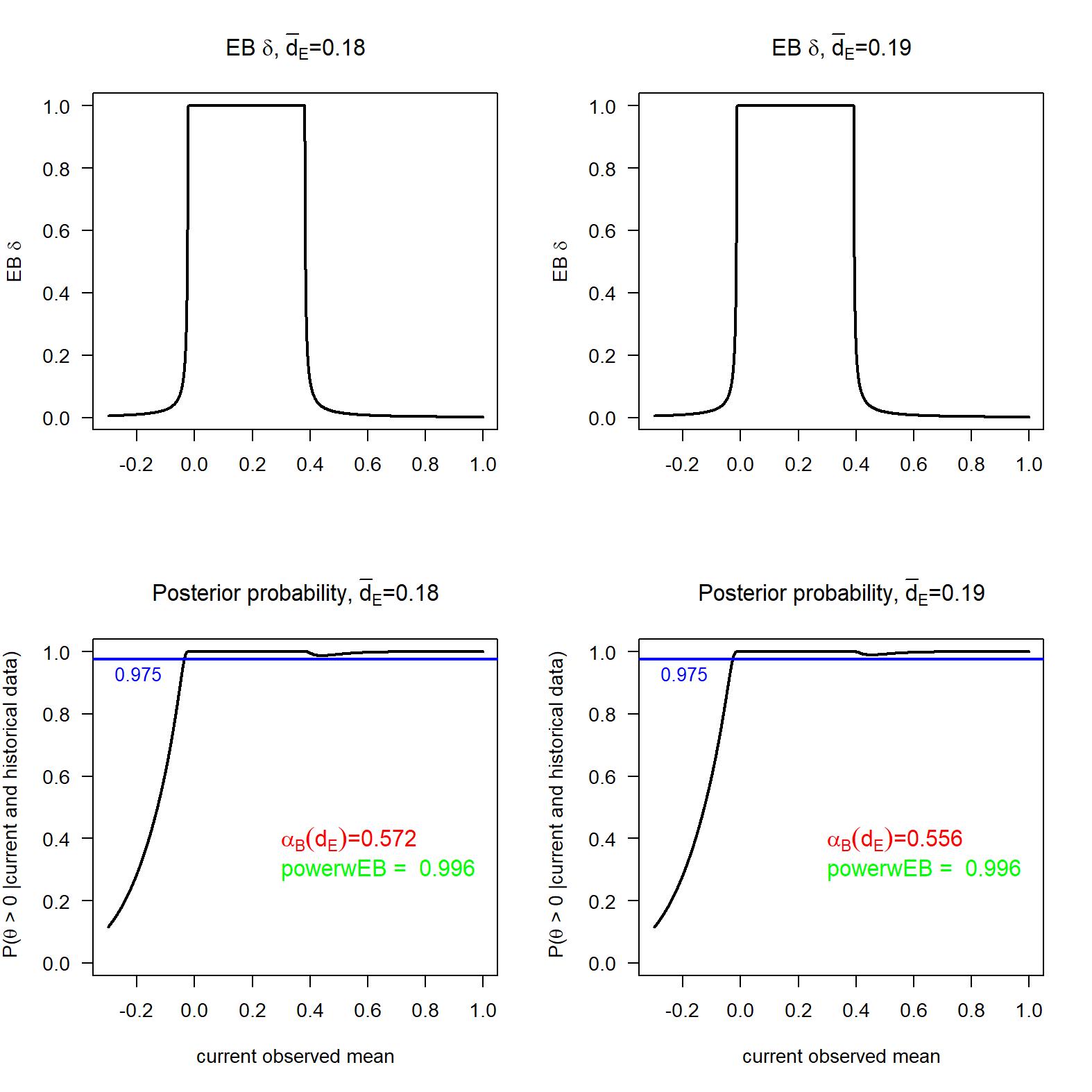}
\caption{$\hat{\delta}(d;d_E)$ (top) and the posterior probability $\Prob(\theta > \theta_0 \given  D=d; D_E = d_E)$ (bottom) for varying current data mean $\bar d$ for $\bar d_E=0.18$ (left) and  $\bar d_E=0.19$ (right). \label{fig:Extreme_EB9}}
\end{center}
\end{figure}

\begin{figure}[ht!]
\begin{center}
\includegraphics[scale=0.7]{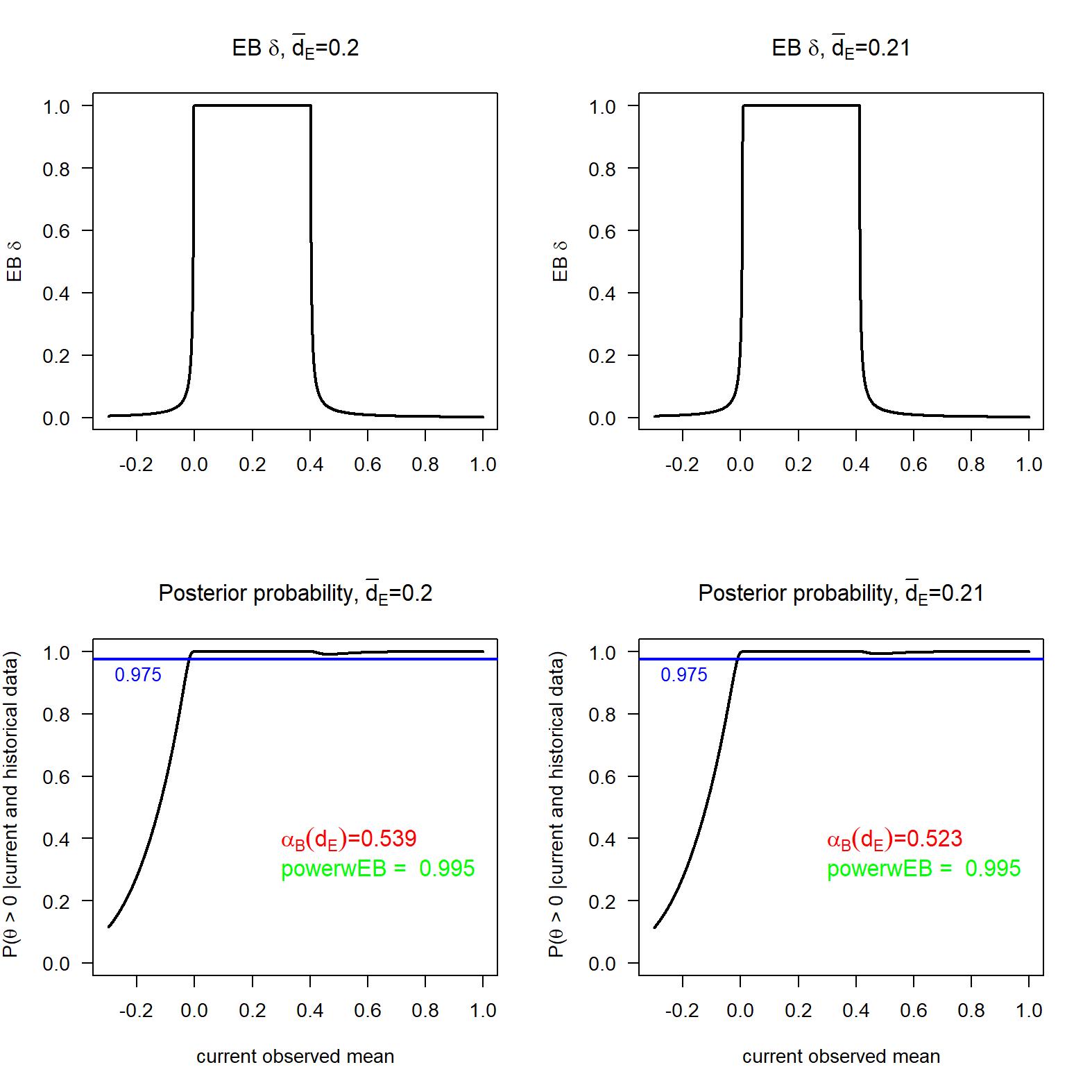}
\caption{$\hat{\delta}(d;d_E)$ (top) and the posterior probability $\Prob(\theta > \theta_0 \given  D=d; D_E = d_E)$ (bottom) for varying current data mean $\bar d$ for $\bar d_E=0.20$ (left) and  $\bar d_E=0.21$ (right). \label{fig:Extreme_EB10}}
\end{center}
\end{figure}

Figures \ref{fig:Extreme_EB0} to \ref{fig:Extreme_EB10} show the estimated Empirical Bayes weight parameter $\hat{\delta}(d;d_E)$ (top) and the posterior probability $\Prob(\theta > 0 \given  D=d; D_E = d_E)$ (bottom) for varying current data mean $\bar d$, borrowing from external data with mean $\bar d_E$ ranging from $0.0$ to $0.21$. Note that these plots are different from Figure 2 of the main paper where the external mean $\bar d_E$ is varying on the horizontal axis. 

The blue line at $0.975$ in the lower plot indicates the separation between the acceptance and the rejection region of the test: for current observed means with posterior probability $\Prob(\theta > 0 \given  D=d; D_E = d_E) \le 0.975$, the tests accepts $H_0$ and it rejects if the posterior probability exceeds $0.975$ (cf. equation (4) in the main manuscript).
Starting at $\bar d_E=0.01$, the posterior probability shows a non-monotone behavior. Up to $\bar d_E=0.05$ and for $\bar d_E \ge 0.15$ , this is without consequence for the rejection region of the test since the non-monotonicity occurs in a range of values with all $\Prob(\theta > 0 \given  D=d; D_E = d_E) \le 0.975$ (for $\bar d_E \le 0.05$) or with all $\Prob(\theta > 0 \given  D=d; D_E = d_E) > 0.975$ (for $\bar d_E \ge 0.15$). For values of $\bar d_E$ between $0.06$ and $0.14$, shown in Figures \ref{fig:Extreme_EB3} to \ref{fig:Extreme_EB7}, the rejection region is no longer a single interval, but separated into two intervals. In comparison to the UMP test for the one-sided one-arm situation (calibrated to borrowing from $\bar d_E$), this leads to a power loss, as observed in Figure 2c in the main manuscript.

Each lower plot also shows the integral over the current data, in red the value of $\alpha(d_E) = \Exp_{\theta =0}[\varphi_{\text{B}}(D;d_E)]$ and in green the power with borrowing, powerwEB $=\Exp_{\theta_1 =0.5}[\varphi_{\text{B}}(D;d_E)]$. These numbers are points on the red and on the green line, respectively, in Figure 2c in the main manuscript.

\
